\documentclass[12pt,preprint]{aastex}

\shorttitle{Titolo}
\shortauthors{Marchesini et al.}

\begin{document}

\title{H$\alpha$ rotation curves: the soft core question}

\author{Danilo Marchesini\altaffilmark{1,2}}
\email{danilom@sissa.it}

\author{Elena D'Onghia\altaffilmark{1,3}}
\email{donghia@mpia-hd.mpg.de}

\author{Guido Chincarini\altaffilmark{4,5}, Claudio Firmani\altaffilmark{5}}
\email{chincarini@merate.mi.astro.it~~firmani@merate.mi.astro.it}

\author{Paolo Conconi\altaffilmark{5}, Emilio Molinari\altaffilmark{5}}
\email{conconi@merate.mi.astro.it~~molinari@merate.mi.astro.it}

\and

\author{Andrea Zacchei\altaffilmark{6}}
\email{zacchei@ts.astro.it}

\altaffiltext{1}{Universit\`a degli Studi di Milano, via Celoria 16, 20133 Milano, Italy}
\altaffiltext{2}{SISSA/ISAS, via Beirut 2-4, 34014 Trieste, Italy}
\altaffiltext{3}{Max Planck Institute for Astronomy, K$\ddot{o}$nigstuhl 17, D-69117 Heidelberg, Germany}
\altaffiltext{4}{Universit\`a degli Studi di Milano-Bicocca, Piazza dell'Ateneo Nuovo 1, 20126 Milano, Italy}
\altaffiltext{5}{Osservatorio Astronomico di Brera-Merate, via Bianchi 46, 23807 Merate (LC), Italy}
\altaffiltext{6}{Osservatorio Astronomico di Trieste, via Tiepolo, 11, 34131
Trieste, Italy}

\begin{abstract}

We present high resolution H$\alpha$ rotation curves of 4 late-type dwarf galaxies and 2 low surface brightness galaxies (LSB) for which accurate HI rotation curves are available from the literature. Observations are carried out at Telescopio Nazionale Galileo (TNG). For LSB F583-1 an innovative dispersing element was used, the Volume Phase Holographic (VPH) with a dispersion of about 0.35 \AA pxl$^{-1}$.

We find good agreement between the H$\alpha$ data and the HI observations and conclude that the HI data for these galaxies suffer very little from beam smearing. We show that the optical rotation curves of these dark matter dominated galaxies are best fitted by the Burkert profile. In the centers of galaxies, where the N-body simulations predict cuspy cores and fast rising rotation curves, our data seem to be in better agreement with the presence of soft cores.

\end{abstract}


\section{INTRODUCTION}

Both in physics and in cosmology we face the fundamental problem of a mass component, dark matter, which we feel confident exists because of dynamical astronomical measurements but which we have not yet detected. The only knowledge we have is that it acts gravitationally and that it dominates preferentially on large scales. It is therefore of particular importance to understand its properties to set guidelines in preparing more focused physics experiments to detect it.

If we wish to explore the properties of dark matter, it is necessary to select dark matter dominated galaxies. In the nuclear regions of high surface brightness galaxies (HSB), the baryonic component (disk and bulge) is dominant and therefore masks the contribution of the dark matter to the total mass distribution. On the contrary, late-type dwarf galaxies provide an excellent laboratory to  investigate the dark halo properties. Due to their large mass-to-light ratio $\Upsilon_{\star}$ and lack of prominent central bulges, these systems are thought to be dark matter dominated at all radii. 

LSBs have blue colors, low metallicities, high gas fractions and very extended disks. The low surface brightness of these galaxies is generally interpreted to be due to their having higher disk angular momenta than HSB. All these properties support the idea that LSBs are unevolved galaxies with low current and past star formation rates (van der Hulst et al. 1993; McGaugh \& de Blok 1997). Therefore these systems  represent ideal candidates for measuring the dark matter distribution and for testing the predictions of theories of galaxy formation.

One of the main goals of current cosmology and particle physics is to determine the nature of the dark matter. The current most popular scenario for the structure formation in the universe is based on the inflationary CDM theory, according to which cosmic structures arise from small Gaussian density fluctuations composed of non-relativistic collisionless particles.

It is well known that while the CDM simulations quite nicely explain most of the observations at large scales, they face serious problems on small scales. One of the most serious discrepancies between theory and observations is related to the central dark density of dark matter dominated galaxies, a fact put forward by \citet{moore94}, \citet{flopri94} and \citet{bu95}. \citet{nfw97} find, from detailed cosmological simulations, that the CDM halos have a singular central density (a cuspy core) while the observations tend to support the evidence of a constant central density (a soft core). Higher resolution N-body simulations have showed that the dark density profile is $\rho \propto$ r$^{-1.5}$ at the center, even more cuspy than the NFW model, and hence in starker contrast to the observations (Moore et al. 1999). \citet{deB96} carried out 21-cm line rotation curves for a sample of LSB galaxies supporting the evidence of a soft core for these systems.

On the other hand, \citet{SwMaTr00} have obtained supplementary data in H$\alpha$ for five LSB galaxies and  argued that the low resolution HI observations suffered from beam smearing which smoothed the rotation curves and smeared out the effects of a strong central mass concentration. \citet{mcGRudB01} and \citet{dBRu01} have analyzed the same data concluding that only one of the five LSB galaxies is really affected by beam smearing. In a forthcoming paper, \citet{vdB01} published a catalog of 20 late-type dwarf galaxies. The problem with these data is that the spatial resolution attained is too poor to distinguish the presence of soft cores. Indeed, as expected, the results of the work are non-discriminating. Their conclusion is that there is no convincing evidence against cuspy cores of dwarf galaxy halos, but they point out that the rotation curves studied are also consistent with the presence of soft cores. However, H$\alpha$ rotation curves of LSB showed in \citet{dBRu01} are in favour of core-dominated halos. 

Results by \citet{SaBu00} and \citet{BoSa01} confirm that most late-type dwarf galaxies are core-dominated in conflict with the cusp predicted by the CDM N-body simulations. Furthermore, an analysis by \citet{Sal01} shows evidence for a soft core also in HSB galaxies.

On galaxy cluster scales, the presence of shallow cores for the dark halos is statistically weak, because not many galaxy clusters have yet been properly observed. On the other hand, strong lensing observations of CL0024+1654 allow the production of a mass map of unprecedented resolution. Using these data, \citet{Ty98} have shown the presence of a soft core, in conflict with the prediction of the NFW model. Recent X-ray data from Chandra for A1795 have revealed that the mass distribution of this cluster is shallow at the center indicating a soft core (Ettori et al. 2001). Firmani et al. (2000, 2001) extended the analysis of the global halo scales (core radius and central density) from HI rotation curves of core-dominated LSB and dwarf galaxies to galaxy clusters with evidence of soft cores, integrating the available information from the literature. In that early work the authors find that the central density is independent on the halo mass and the core radius scales with the mass when the mass range is increased. One of the purposes of this work is to see if, by using a sample of H$\alpha$ rotation curves, we are in agreement with their previous findings.

This paper is organized as follows: in Section 2 we describe our observations and the data reduction while in Section 3 we present our rotation curves and compare our observations with earlier HI data. The mass model and the fits of the observations are presented in Section 4 with the global halo scaling relations. Our results are briefly summarized in the Section 5. We use $H_{o}$=75 km s$^{-1}$ Mpc$^{-1}$ throughout the paper.

\section{THE DATA}

\subsection{Sample}

The galaxies presented in this work have been selected from the sample of late-type dwarfs in \citet{vdB01}, and from the sample of LSB galaxies in \citet{deB97}. The properties of these galaxies are presented in Table~\ref{tbl-1}, which lists the galaxy name (1), the adopted distance using $H_{0}$=75 km s$^{-1}$ Mpc$^{-1}$ (2), the disk scale length in kpc (3), the inclination angle (4), the position angle (5), the systemic velocity (6), the central surface brightness (7), the morphological type (8) and the references (9). The first four galaxies are late-type and dwarf galaxies, while the last two are LSB galaxies.

\subsection{Observations}

Observations were carried out at the 3.6~m Telescopio Nazionale Galileo (TNG)
in November 2000 and March 2001. We used the {\it d.o.lo.res.} instrument in
spectroscopic mode with the grism HR-r (0.8 \AA~pxl$^{-1}$, $\lambda \in$
(6200,7800) \AA~and a spatial scale of 0.28$^{\prime \prime}$pxl$^{-1}$) for
all the galaxies except F583-1. For this galaxy we used an innovative VPH
(Vo\-lu\-me Phase Holographic) dispersing element with 1435~lines/mm and
$\lambda_{blaze}=\lambda_{H\alpha}$ (the dispersion was
about 0.35 \AA~pxl$^{-1}$). The slit size was either 1$^{\prime \prime}$ or
1.5$^{\prime \prime}$ depending on the seeing, yielding a spectral resolution
of $\sim$ 3 \AA~(4.5 \AA). We used {\it d.o.lo.res.} in photometric mode for the
positioning of the slit: by doing this, we were able to ensure that the slit
was properly aligned along the major axis of the galaxy as seen on the screen. Cumulative exposure
times were half an hour, except for UGC4325 and UGC7603 for which they were
one hour. All the 6 ga\-la\-xies had previous HI published observations. For all galaxies except
UGC4325 and UGC4499 Argon comparison lamp frames were taken immediately before
and after each object exposure; for UGC4325 and UGC4499 this was not possible
because of technical problems, but this did not significantly affect the resulting rotation curves.

\subsection{Reductions}

Red spectra of spiral galaxies at low redshifts typically show the prominent
nebular lines of H$\alpha$ $\lambda$6563, [N II] $\lambda$$\lambda$6548,6583,
and [S II] $\lambda$$\lambda$6716,6731; usually, the disk emission is stronger
in H$\alpha$ than in [N II] and [S II]; the H$\alpha$ line is also more spatially
extended than the other four emission features in our six
galaxies. For these reasons, we will focus only on the H$\alpha$ rotation
curves. The strength of the H$\alpha$ line, and hence the accuracy of the rotation curves, varies from object to object.

\subsubsection{Preliminary Reductions}

Standard calibration material includes high signal-to-noise flats (with same slit opening as objects), a series of bias frames and Argon comparison lamp frames. Standard reduction steps were taken: bias subtraction, flat-fielding, and removal of cosmetic defects (hot pixels, bad columns and cosmic rays) were all done using procedures in IRAF.

\subsubsection{Wavelength Calibration and Sky Subtraction}

A polynomial pixel-wavelength fit was extracted from each Ar calibration spectrum using interactive line identification. Eight to eleven lines were identified in each of the lamp spectra, and a third-order polynomial was fit to derive the di\-spersion curve. The same calibration lamps were used to model the effect of line-curvature. For UGC4325 and UGC4499, for which we could not take Ar comparison lamp frames right before and after the object spectrum, we used the sky lines for wavelength calibration.

We checked the goodness of the wavelength calibration on the sky lines for each object spectra; that is, after  calibration, sky lines should be straight (along the spatial direction) and match the laboratory wavelengths. For all the galaxies except F583-1 the one sigma calibration velocity error was $\delta v_{cal}$ $\in$ (4,7) km s$^{-1}$, while for the galaxy F583-1 $\delta v_{cal} \approx$ 2.5 km s$^{-1}$.

After the wavelength calibration test, the sky lines and the sky background  were removed by interpolating the regions above and below the galaxy emission with a second-order polynomial.

\begin{figure}
\caption{Example of a two dimensional long slit spectrum after reduction: LSB F583-1\label{f1.ps}}
\vskip 0.5truecm
\epsscale{0.8}
\plotone{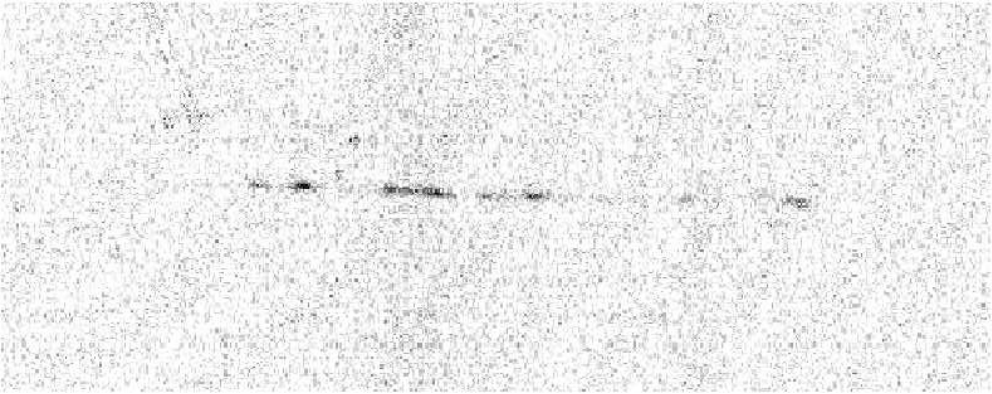}
\end{figure}

\begin{figure}
\caption{Image of the galaxy LSB F583-1 with the exact position of the long slit\label{f2.ps}}
\epsscale{0.8}
\plotone{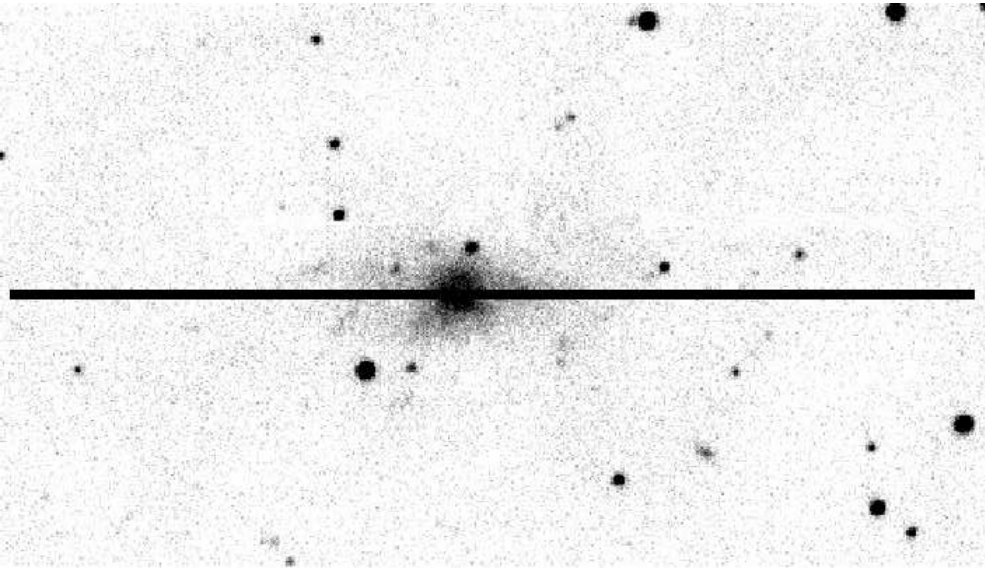}
\end{figure}

An example of a two dimensional long slit spectrum after reduction is shown in Figure~\ref{f1.ps} for the galaxy LSB F583-1; Figure~\ref{f2.ps} shows the exact position of the long slit on the galaxy.

\subsubsection{The Rotation Curves}

We measured the centroids at each position (row) along the H$\alpha$ galaxy emission by making Gaussian fits to the line profile. The errors were derived by simulating realistic spectra, and estimating the relation $\delta v_{S/N}=F(S/N)$ for each spectrum; the S/N threshold was chosen such that $\delta v_{S/N} \lesssim$ 25 km s$^{-1}$. The final one-sigma velocity error at each position along the major axis is therefore given by $\delta v = \sqrt{(\delta v_{cal})^{2}+(\delta v_{S/N})^{2}}$. 

We assumed, furthermore, that the maximum of the light in the direction perpendicular to the dispersion of the continuum spectrum would coincide with the center of the galaxy. However, in those cases where the continuum was too weak to give a reliable estimate, we used a sigmoid function to fit the unfolded rotation curve and assumed the center of symmetry to be the center of the galaxy. In other words we got an estimate of the dynamical center under the assumption of unperturbed circular motion. In all cases the fit was carried out accounting for the errors estimated as above and the velocity of the center of symmetry was assumed to be the systemic velocity of the galaxy.

The final rotation curves (velocity as a fun\-ction of the galactocentric distance) were obtained by folding the line-of-sight velocity profiles around the estimated centers (i.e. by combining the approaching and the receding sides) after correction for the inclination of the galaxies ($v(r)=v_{l.o.s.}/\sin{i}$), and by re-sampling the H$\alpha$ points every 1.5$^{\prime \prime} \div$3$^{\prime \prime}$, depending on the seeing and the H$\alpha$ emission distribution (continuous or clumpy) of each galaxy.

The final one-sigma error $\Delta v$ for each radial velocity point also takes into account the asymmetries in the rotation curve, which are less than 10 km s$^{-1}$. Finally $\Delta v$ is given by:
\begin{equation}
\Delta v=\sqrt{rms_{v}^{2}+\frac{1}{\sum_{i=1}^{N}{(1/\delta v_{i}^{2})}}},
\end{equation}
where $rms_{v}$ is the standard deviation of the $N$ velocity measurements inside the re-sampling bin and $\delta v_{i}$ is the one-sigma error of the line-of-sight velocity measurements in the bin. The $\gtrsim$ 50 \% asymmetries in the rotation curve (quite common in late-type dwarf galaxies as shown by Richter \& Sancisi 1994) are taken into account in the term $rms_{v}$.

The final one-sigma errors for the velocities are, as expected, fairly large when compared to the HI observations, except for the
two galaxies UGC7603 and F583-1. For all the galaxies, the major contribution
to the final error comes from the $\delta v_{S/N}$ term: because of
the short exposure times and the intrinsic faintness of the H$\alpha$ emission in these galaxies, the H$\alpha$ line did not have a high enough S/N to make the term $\delta v_{S/N}$ small. The rotation curve of UGC7603 has instead quite small errors, due both to the higher cumulative exposure time and to intrinsicly stronger H$\alpha$ emission. The rotation curve of the galaxy F583-1 has very low velocity errors, due to the fact that the corresponding spectrum was taken with the new VPH di\-spersing element. This newly introduced upgrade in the {\it d.o.lo.res.} spectrograph allows a higher spectral resolution together with an enhanced transmission efficiency (Conconi et al. 2001).

\section{COMPARISON WITH OTHER OPTICAL ROTATION CURVES}

\begin{figure}
\caption{Comparison between our H$\alpha$ and the H$\alpha$ rotation curves in 
\citet{dBRu01}: open squares are our H$\alpha$ points; asterisks are the
H$\alpha$ points of the rotation curves in \citet{dBRu01}; the error bars
shown represent one-sigma errors\label{f3.psf4.ps}}
\vskip 0.5truecm
\epsscale{1.0}
\plottwo{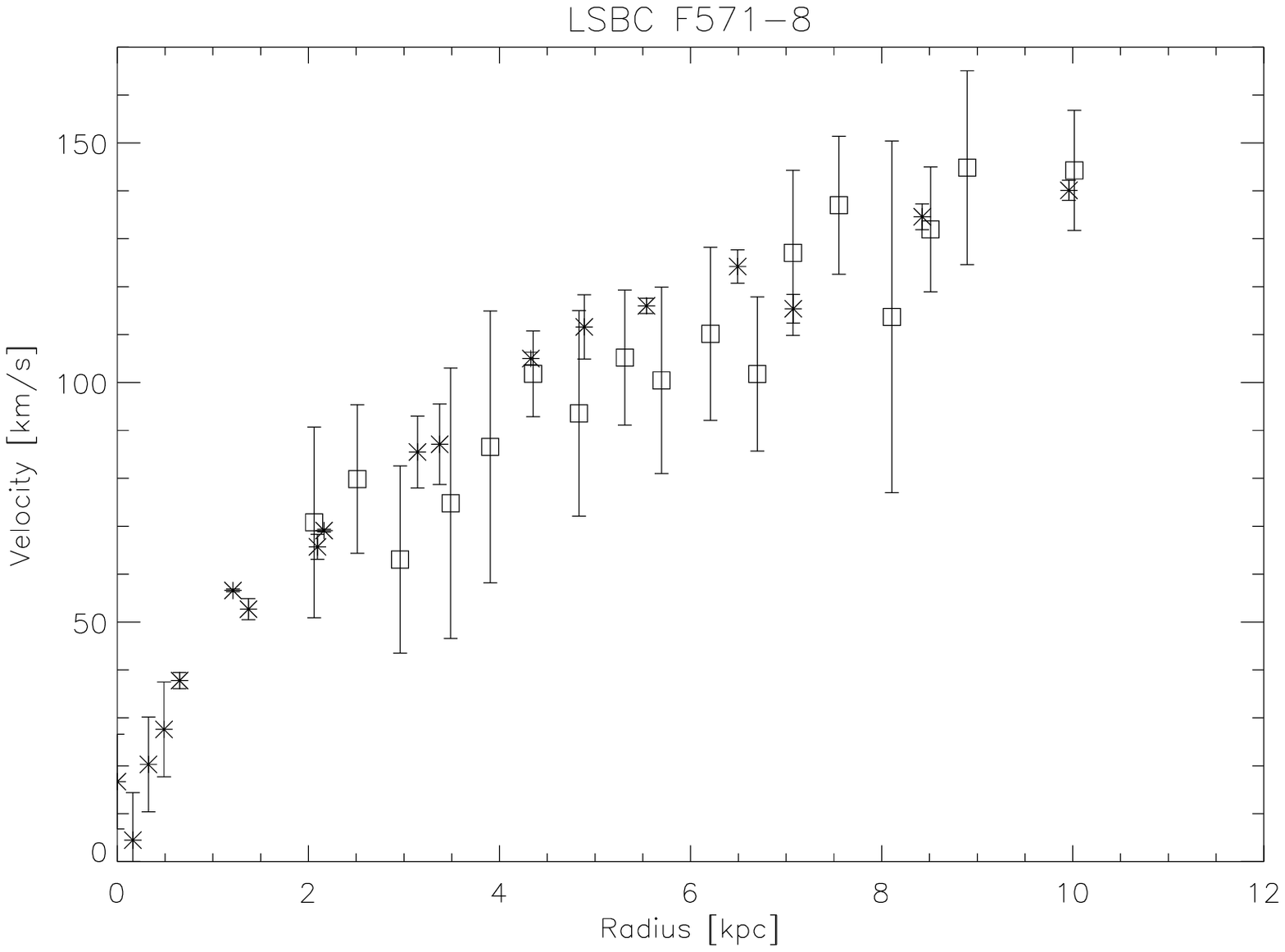}{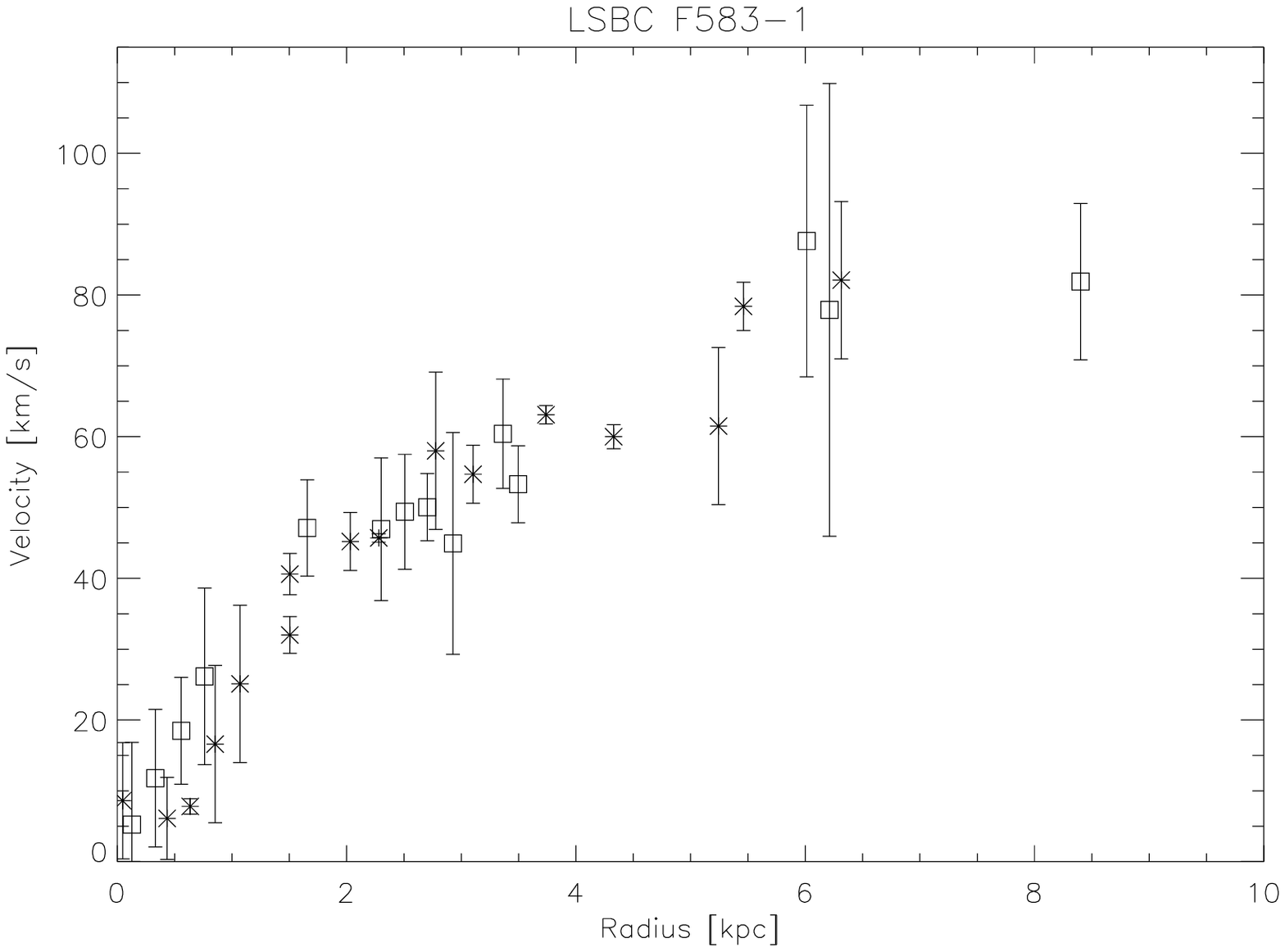}
\end{figure}

High resolution H$\alpha$ rotation curves for
the two galaxies F571-8 and F583-1 have been published by \citet{dBRu01}. In Figure~\ref{f3.psf4.ps} our H$\alpha$
rotation curves for these two galaxies are plotted together with the H$\alpha$
rotation curves obtained by \citet{dBRu01}: for both galaxies, the two
independent rotation curves agree very well within the errors. 

\subsection{H$\alpha$ vs HI Data}

\begin{figure}
\caption{Comparison between H$\alpha$ and HI rotation curves: asterisks are
the H$\alpha$ points; open squares are the HI points; for each galaxy is also
specified the spatial resolution of the rotation curve in arcsec, both for
H$\alpha$ and for HI observations; the error bars shown represent one-sigma errors\label{f5.psf6.psf7.psf8.psf9.psf10.ps}}
\vskip 0.5truecm
\plottwo{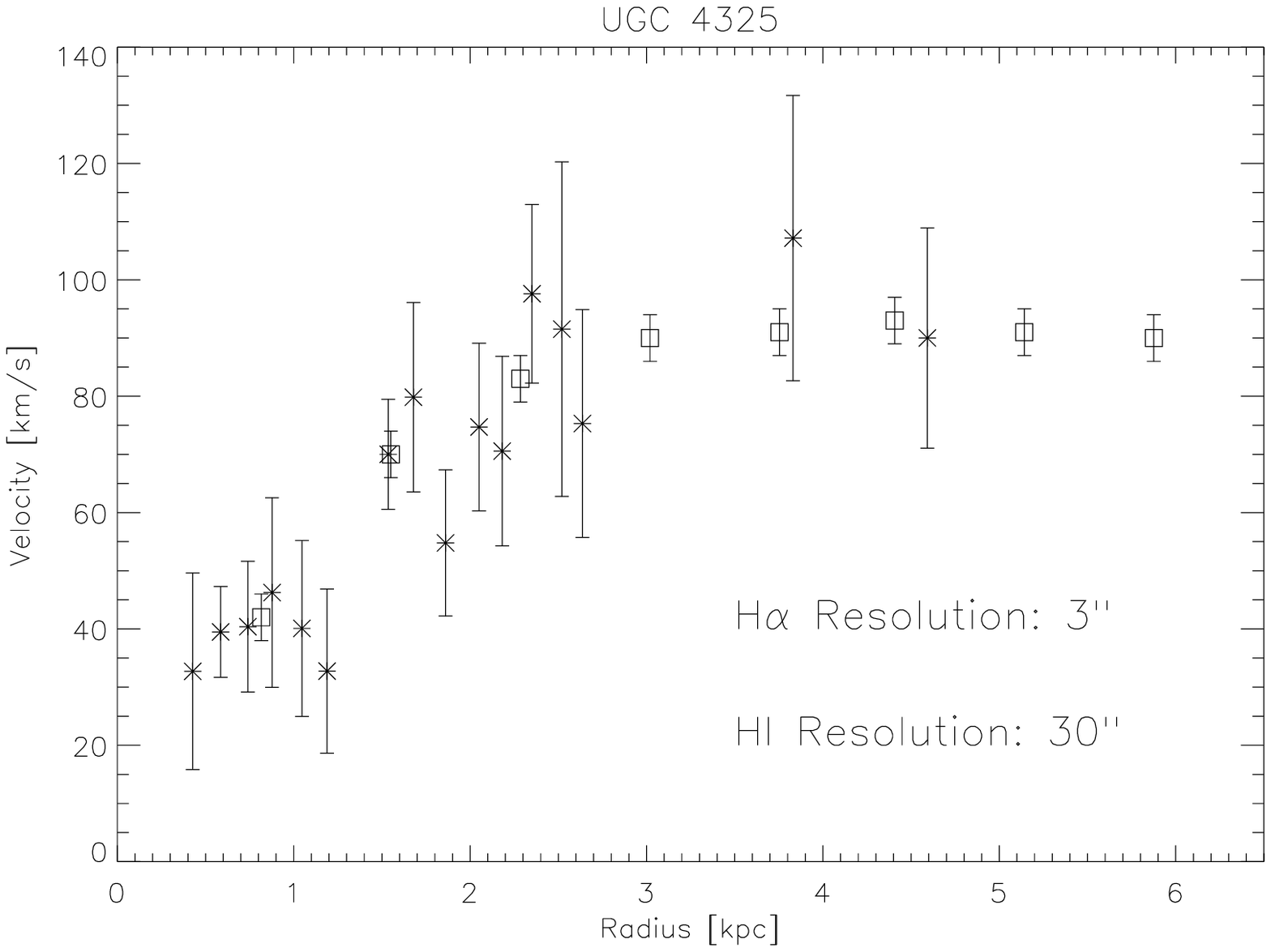}{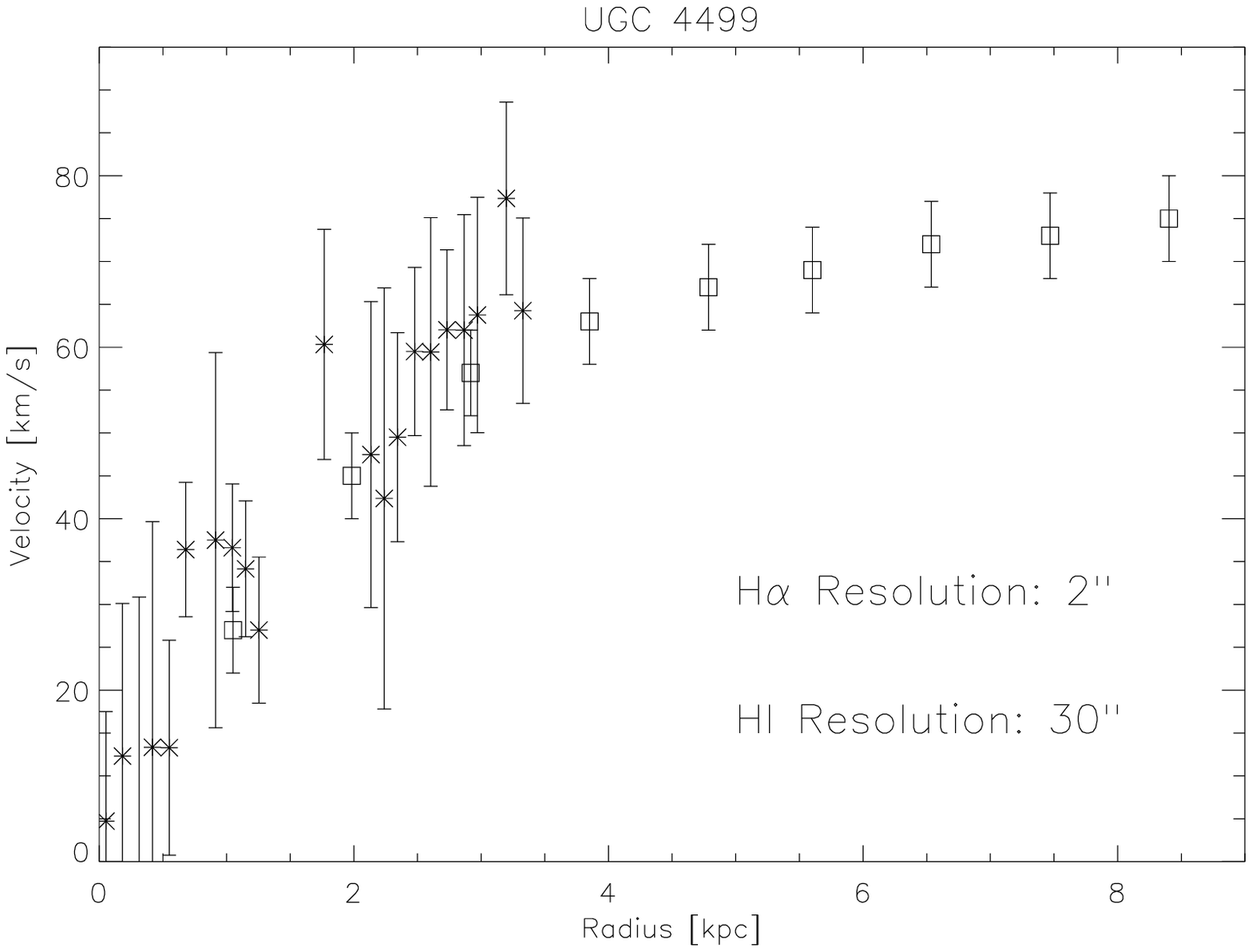}
\end{figure}

The observed H$\alpha$ rotation curves were then compared to the HI rotation curves published in literature, in order to see how strongly the HI rotation curves are affected by beam smearing and to complement the H$\alpha$ curves at large distance from the center. Indeed the spatial resolution of the HI data for the LSB ga\-la\-xies (F571-8 and F583-1) is $\gtrsim$13$^{\prime \prime}$, while the spatial resolution of the HI data for the late-type dwarf galaxies is $\approx$30$^{\prime \prime}$. In Figure~\ref{f5.psf6.psf7.psf8.psf9.psf10.ps} we show the H$\alpha$ rotation curves (this work) with the 21 cm points overplotted (de Blok \& McGaugh 1997; van den Bosch \& Swaters 2001).

For the three galaxies UGC4325, UGC7603 and UGC11861 the H$\alpha$ points agree (within measurement errors) with the HI data; for the two galaxies UGC4499 and F583-1 the H$\alpha$ velocities are slightly larger than the HI data in the inner regions, but the velocity differences are always $\lesssim$ 5-10 km s$^{-1}$; only for F571-8 are the H$\alpha$ velocities much larger (up to 30 km s$^{-1}$) than those derived from the 21 cm line.

We conclude that for five of the six galaxies in our sample, the HI rotation curves
are not affected substantially by beam smearing. The only galaxy
with an HI rotation curve which is apparently affected by beam smearing is F571-8. This galaxy is almost edge-on so the discrepancies could be due to differences in the optical depth and projection effects.

\section{MODELING}

For robust mass modeling, both good spatial resolution in the inner regions and extension of the rotation curves to large radii are necessary. The good spatial resolution in the inner regions is necessary to properly estimate the mass distribution in the nuclear region and to discriminate between cuspy or soft dark matter density cores. The extension to large radii is necessary to better constrain the total mass of the galaxy; it also helps costrain the NFW fits, whose velocities tend to keep rising if not constrained at large radii.

Because the H$\alpha$ curves have good spatial resolution and the 21 cm curves go to larger radii, we combine the data to produce hybrid rotation curves. We use these hybrid rotation curves in the following mass modeling in order to see if the observed rotation curves are characterized by soft or cuspy density profiles.

The aim of this work is to test for the existence of soft cores for the observed galaxies. We want to give to the NFW model a maximum chance to match the data. Thus, since for these galaxies the luminous component contribution to the gravitational potential seem not to be dominant, we assume that the observed optical high-spatial resolution rotation curves trace the dark halo component. We neglect, in a first approximation,  the disk contribution to the total gravitational potential.  The idea is to verify if the observed rotation curves rise more gently in the inner regions than the NFW model, showing evidence of a soft core. A more accurate modeling of the rotation curve makes the conflict with the CDM models even worse.

In this approximation we apply a best fit procedure to the data assuming three different models for the dark halo mass distribution and consequently three different functions for the circular velocity:
\begin{itemize}
\item the Burkert profile (Burkert 1995):
\begin{equation}
v^{2}(r) = 2 \pi G \rho_{o} r_{o}^{3} \frac{1}{r} \Bigg \{ \ln{\Bigg[\bigg(1 + \frac{r}{r_{o}}\bigg) \sqrt{1 + \bigg(\frac{r}{r_{o}}\bigg)^{2}} \Bigg]} - \arctan{\frac{r}{r_{o}}} \Bigg \},
\end{equation}
where $\rho_{o}$ and $r_{o}$ are respectively the central density and the core radius; this profile is characterized by a central soft density core ($\rho$ is constant  at small radii), while at large radii $\rho \propto$ r$^{-3}$.
\item the NFW profile (Navarro, Frenk \& White 1997):
\begin{equation}
v^{2}(r) = 4 \pi G \rho_{o} r_{s}^{3} \frac{1}{r} \Bigg[\ln{\bigg(1 + \frac{r}{r_{s}} \bigg)} - \frac{r/r_{s}}{(r/r_{s}) + 1}\Bigg],
\end{equation}
where $\rho_{o}$=$\delta_{c} \rho_{crit}$ and $r_{s}$ is a scale radius; this profile is characterized by a cuspy core ($\rho \propto$ r$^{-1}$ at small radii), while at large radii $\rho \propto$ r$^{-3}$.
\item the Moore profile (Moore et al. 1999):
\begin{equation}
v^{2}(r) = \frac{8}{3} \pi G \rho_{o} r_{s}^{3} \frac{1}{r} \ln{\Bigg[1 + \bigg(\frac{r}{r_{s}} \bigg)^{1.5} \Bigg]},
\end{equation}
where $\rho_{o}$ is a characteristic density and $r_{s}$ is a scale radius; this profile is characterized by a steeper central cuspy density core ($\rho \propto$ r$^{-1.5}$ at small radii), while at large radii $\rho \propto$ r$^{-3}$.
\end{itemize}

The observed rotation curves were normalized to the outermost observed points; we also normalized the three models to the corresponding maximum velocities and maximum radii ($r_{max}=3.3r_{o}$ for the Burkert profile, $r_{max} \approx 2.16r_{s}$ for the NFW profile, and $r_{max} \approx 1.25r_{s}$ for the Moore profile).

\begin{figure}
\caption{Normalized hybrid rotation curves compared to models in minimum disk hypothesis: asterisks are the H$\alpha$ points, open squares are the HI data; the continuous line is the Burkert profile, the dashed line is the NFW profile, and the dotted-dashed line is the Moore profile; the observed rotation curves  are normalized to the last observed point; the numbers in parenthesis are the normalized chi-squared $\mathrm{\chi}_{\nu}^{2}$ for the three models; the error bars shown represent one-sigma errors.\label{f11.psf12.psf13.psf14.psf15.psf16.ps}}
\vskip 0.5truecm
\plottwo{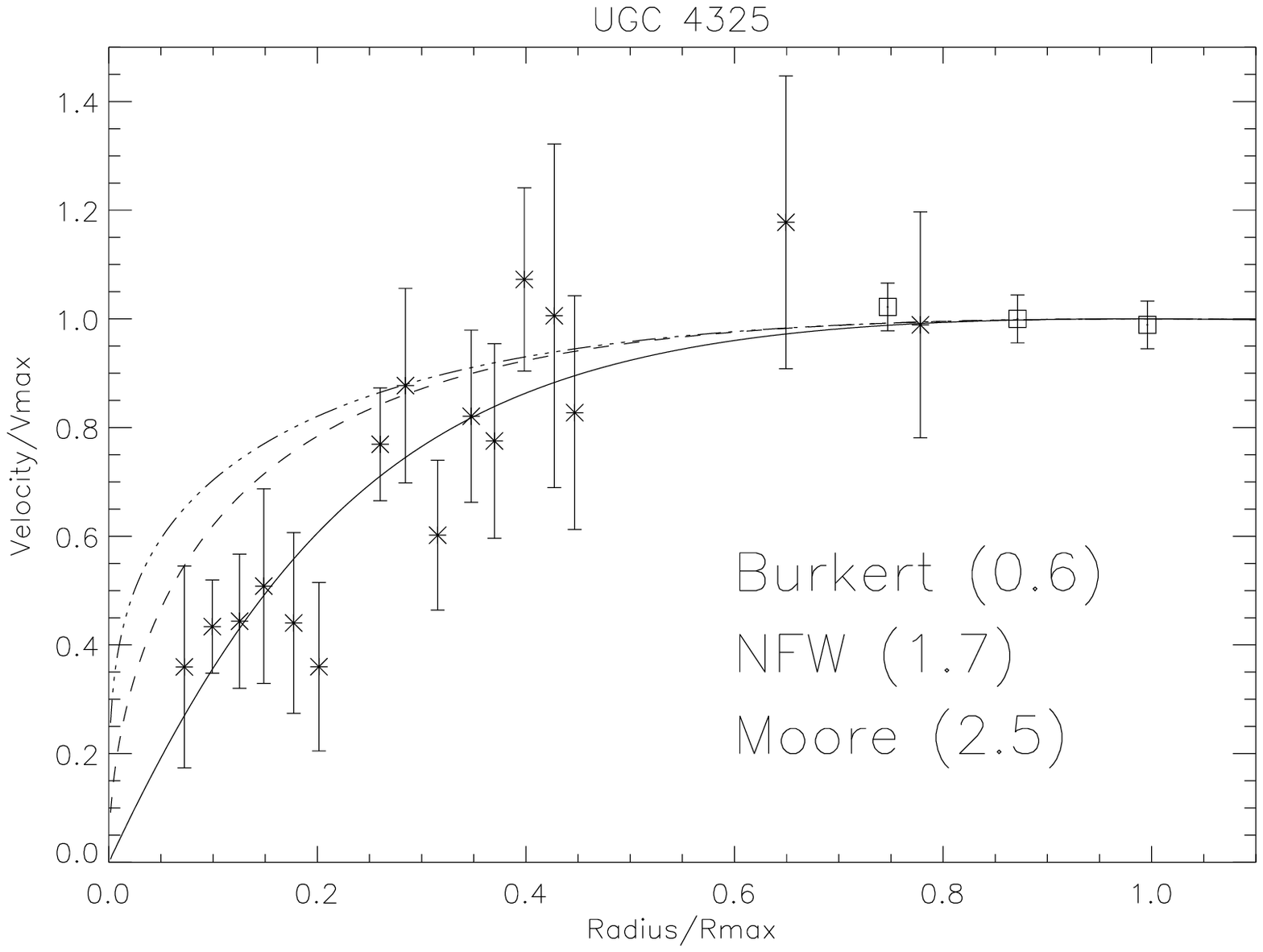}{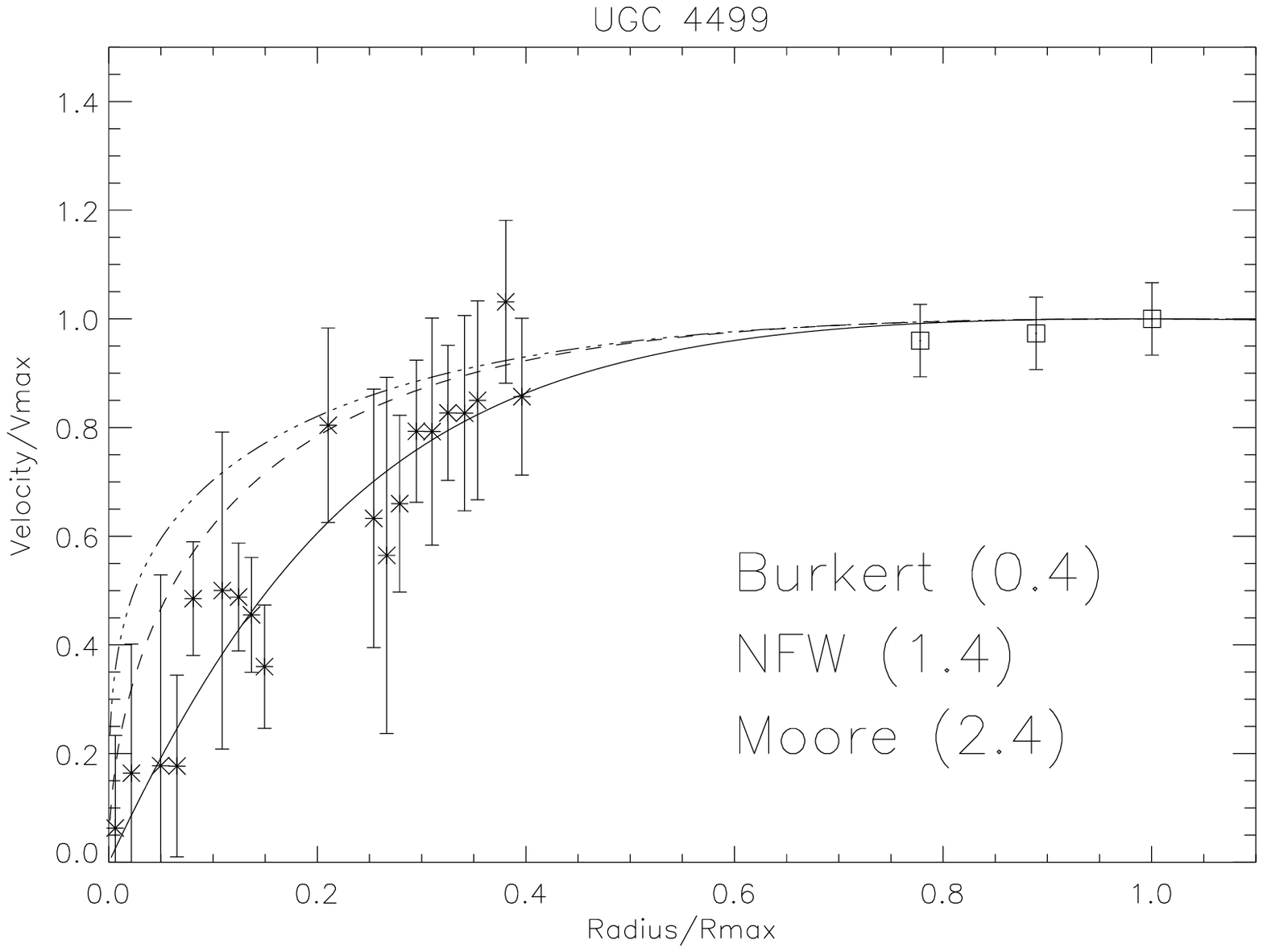}
\end{figure}

The observed rotation curves are shown in Figure~\ref{f11.psf12.psf13.psf14.psf15.psf16.ps} overplotted with the three models. 
First of all, we notice how nicely the Burkert
profile follows the observed rotation curves (we must keep in mind that we
did not fit the observed points, but we only normalized each curve to its last point). Secondly, the NFW and the Moore profiles are inconsistent with the observed rotation curves in the inner regions: they both predict a too fast rising rotation curve because of the presence of the cuspy cores. The natural matching of the Burkert profile to the observed rotation curves and the inconsistency in the inner regions seen when we use the two profiles predicted by the numerical simulations tell us that these six galaxies seem to be core dominated. Thirdly, the Moore profile is always the most inconsistent in the inner points, since considering steeper inner density profiles only makes the inconsistency greater between the observed and the theoretical rotation curves. 
In this first fit procedure, note that the NFW model and the Moore profile have normalized chi-squared with large values, while low values are only obtained assuming the Burkert profile. In the plot, the normalized chi-squared are indicated in parenthesis for the three models, respectively.

A second comparison with the NFW profile for the mass distribution of the galaxies may be obtained. The NFW circular velocity as a function of the radius may be written in terms of the concentration parameter c=$r_{200}/r_s$ ($r_{200}$ is the radius of the sphere having mean density equal to 200$\rho_{crit}$) and the circular velocity at the virial radius $v_{200}$:

\begin{equation}
v^{2}(r) = v_{200}^{2} \frac{1}{x} \frac{\ln{(1+cx)} - [cx/(1+cx)]}{\ln{(1+c)} - [c/(1+c)]},
\end{equation}
where $c$ and $v_{200}$ are the two free parameters of the fit and $x=r/r_{200}$.

The best fit procedure allows us to derive the values of c and $v_{200}$ of the observed rotation curves.
The fitting is done by the maximum-likelihood method applied to the function:
\begin{equation}
\chi^{2}_{\nu}=\frac{1}{N-m} \sum_{i=1}^{N}{\frac{1}{{\delta v_{i}}^{2}} [v_{i} - v(r_{i})]^{2}},
\end{equation}
where $N$ is the total number of points to fit, $m$ is the number of free parameters, $\delta v_{i}$ is the one-sigma velocity error, $v_{i}$ is the observed velocity and $v(r_{i})$ is the model velocity.

In Table~\ref{tbl-3}~and~\ref{tbl-4} the minimum disk model best fit scales are shown respectively for the Burkert and the NFW profiles.

\begin{figure}
\caption{Theoretical $c-M_{200}$ relation (NFW model in a $\Lambda$CDM universe) with overplotted the observed points from the minimum disk analysis of the six galaxies; the error bars represent the one-sigma errors.\label{f17.ps}}
\vskip 0.5truecm
\plotone{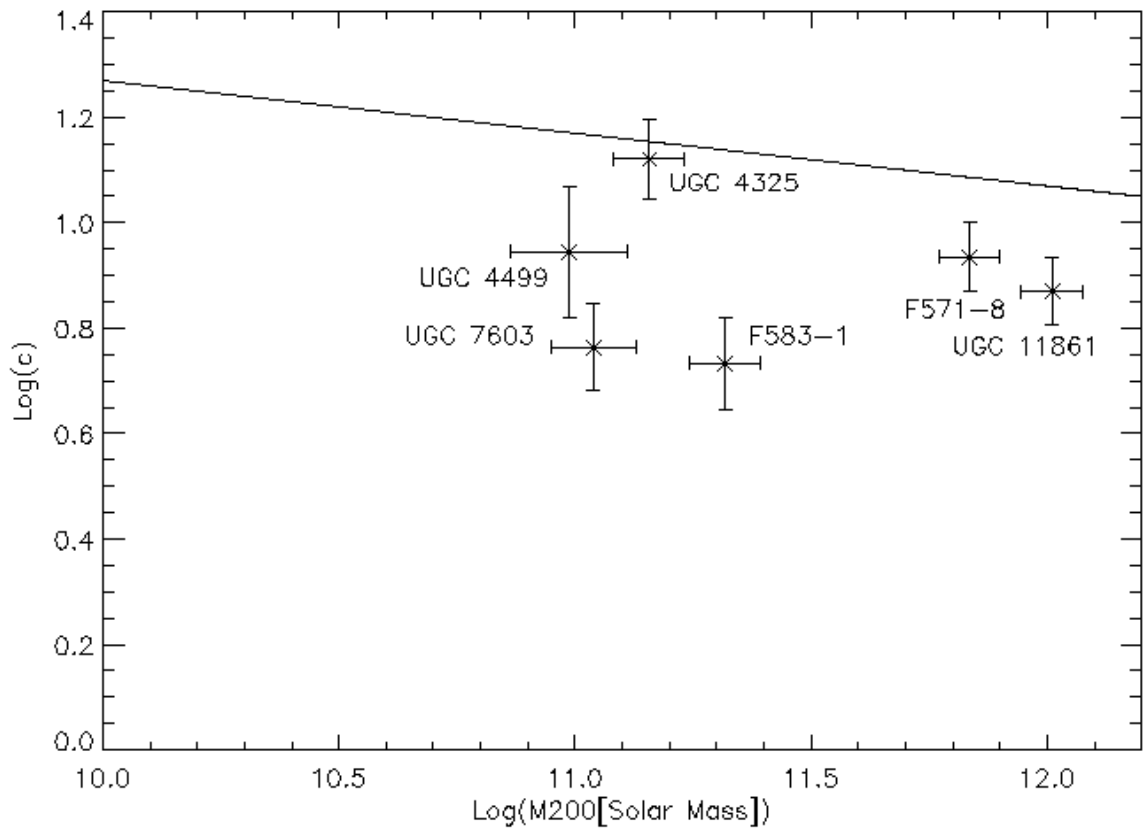}
\end{figure}

The Burkert profile reproduces both the inner and the
outer regions of the observed rotation curves, while the NFW profile best fit
solution does not reproduce the rotation curves as well. To obtain a good
enough statistical fit with the NFW profile, we had to force the fitting
procedure; we found that the best fit solutions are characterized by too large maximum velocities, in that the best fitting maximum velocities are up to 30$\%$ higher than what is observed. Note that the outermost points of the rotation curves come from HI 
observations taken from the literature with errors
of $\sim$ 3-7 km s$^{-1}$. The galaxy UGC4499 is the only one for which the
NFW best fit solution does not require a large maximum velocity, but this
galaxy has a stellar bulge component which, even if not dominant, could make  the inner profile steeper. The concentration values derived by the best fit procedure are listed in the Table 4 for the observed objects; the observed $c-M_{200}$ relation is shown in Figure~\ref{f17.ps} overplotted with the theoretical one predicted by the NFW model for a $\Lambda$CDM cosmology. Note that even forcing the NFW model to represent the mass distribution of the galaxies, the concentrations appear to be lower than the predicted values from the CDM N-body simulations. Unfortunately, the concentration values measured in the dark halos by the N-body simulations show a very large scatter  and no consensus is reached by different authors on the real scatter in the relation concentration-halo virial mass (or c-$M_{200}$) (see e.g. Wechsler 2001). Note however that the values inferred for the concentrations are upper limits in this minimum disk approximation. Thus, if the disk component is included in the analysis, the concentration values should be pushed down to lower values in the plot.

Regarding a comparison with other published works, the galaxies F583-1 and F571-8 are also found in the published sample of high resolution H$\alpha$ rotation curves by \citet{dBRu01}. For the LSB F583-1 we have a good agreement between our estimate of the central density assuming a Burkert profile as representative of the mass distribution and the minimum disk model adopted in \citet{dBRu01} analysis. Since the core radius estimate depends on the assumed model, and the \citet{dBRu01} analysis uses a isothermal model, an agreement is not necessary required. For F571-8 our estimate of the central density in this approximation is slightly lower than the \citet{dBRu01} estimate. Because of the projection effects associated with our seeing this galaxy edge-on, we conclude that our estimates of the central densities are close to the \citet{dBRu01} work. The two galaxies UGC4325 and UGC7603 are in common with the work of \citet{dBBo02}. For UGC4325 there is disagreement between our estimates of $c$ and $v_{200}$ and the corresponding ones obtained by \citet{dBBo02}; they find a lower value for $c$ and a higher value for $v_{200}$ than we do. For this galaxy we also find a value of $\rho_{o}$ for the Burkert profile which is more than twice the corresponding value found by \citet{dBBo02}. The two parameters $\rho_{o}$ and $r_{o}$ are related in the way that if $r_{o}$ is increased, $\rho_{o}$ must decrease in order to fit a given rotation curve; although $\rho_{o}$=0.231 $M_{\odot}$ pc$^{-3}$ and $r_{o}$=1.77 kpc are the best fit parameters for our H$\alpha$ rotation curve, it may be possible that a better rotation curve (with smaller velocity error bars and with more points in the inner region) could better constrain the two parameters, giving a larger value for the core radius and a smaller value for the central density. For UGC7603, instead, there is quite good agreement on all the best fit parameters for the two assumed minimum disk models.

We therefore conclude that the six observed galaxies, in the minimum disk
hypothesis, are characterized by dark matter profiles with constant density
cores and are rather inconsistent with the density profiles predicted by cosmological
numerical simulations. This inconsistency is worse in the case of the Moore profile. By taking into account
both the contribution of the stellar disk and the baryonic adiabatic contraction, the inconsistency between the NFW profile and the observed rotation curves gets worse.

\subsection{Halo Adiabatic Contraction}

In the last section we found that the galaxies in our sample are core dominated. However, the values for the halo central density and the core radius were inferred by a first analysis, neglecting the disk component. They have to be considered as upper limits for the halo scales and are representative of the present dark halo. In the following section, we analyze the rotation curves taking into account the disk contribution.  The dark matter halo density profile that we infer today from observations is not the primordial one. In fact, due to the cooling of the baryons inside the virialized halo, the halo gravitational potential well is altered and the halo shrinks, becoming more concentrated in the inner regions, and forcing the matter distribution to re-adjust. Consequently, after the disk formation, the dark component of the rotation curve could rise more steeply in the inner regions. If our aim is to test the predictions of the current cosmological paradigm through the mass distribution of the observed galaxies, we have to take into account the halo contraction during the disk formation and correct our analysis by this effect.

 Since the disk formation is a slow process, the halo contraction is assumed to be adiabatic and we account for it by using adiabatic invariant techniques (Flores et al. 1993). A useful approximation is to assume that the angular momentum of the dark matter particles is unaffected by the baryons that are collapsing toward the center. That is, for circular motion, $r~M(<r)$ is preserved, and this quantity is an adiabatic invariant during the growth of the disk. In this way, it is straightforward to predict a rotation curve reflecting the present distribution of dark matter and baryons from an initial halo profile. Accounting for the halo adiabatic contraction we can attempt to discriminate between the NFW and the King model as representative mass distribution of the primordial dark halo. 

The procedure we have implemented in order to interpret the data builds a fiducial galaxy. We assume for the primordial halo a mass distribution with a soft core (described by a King model) or a cuspy core (assuming the NFW profile). The King model is assumed with a form parameter $f_{p}=8$. Although the shape of the profile in the central region is not sensitive to the form parameter, we chose to use this particular King model, since the rotation curves of LSB and dwarf galaxies are well fitted by this model in the inner regions (Firmani et al. 2001).

We build the disk in order to reproduce the observed photometric data. We have our own photometric data only for UGC4325 and UGC4499. For the other galaxies, the photometric values of LSB F583-1 and LSB F571-8 are taken from \citet{dBRu01} and of  UGC11861 and UGC7603 from \citet{Swa99}. In Table~\ref{tbl-5} the baryonic fraction $f_b=M_{disk}/M_{vir}$ for the observed galaxies is reported with the assumed mass to light ratio $\Upsilon_{\star}$ and the scale length $h$. The disk was modeled using the photometric profile with $\Upsilon_{\star}$=1.4 for the two LSB galaxies (consistent with the E(B-V)=0.6 estimated for these galaxies by de Blok et al. 2001) and $\Upsilon_{\star}$=1.0 for the late-type dwarf galaxies. This value was estimated using the model provided by \citet{ChLo01} with a star formation rate $SFR \propto e^{-t/\tau}$ ($t \approx$ 7 Gyr and $\tau \approx$ 15 Gyr)(Longhetti, private  communication). Then, we correct for the adiabatic contraction using the adiabatic invariant technique and we derive the final rotation curve to be compared to the data.

\begin{figure}
\caption{Comparison between the observed rotation curves and the models in which the adiabatic contraction and the stellar disk component are taken into account; the left plots correspond to the models in which the primeval dark matter halo density profile is represented by a NFW model, while the right plots correspond to the models in which the primeval dark matter density profile is represented by a King model. Asterisks are the H$\alpha$ points, open squares
are the HI points; the dotted-dashed line is the contribution of the stellar
disk, the dashed line is the primeval dark matter halo velocity profile, and
the continuous line is the final rotation curve (after taking care of the
adiabatic contraction) which must be compared to the observed rotation
curve.\label{f18.psf19.psf20.psf21.psf22.psf23.psf24.psf25.psf26.psf27.psf28.psf29.ps}}
\vskip 0.5truecm
\plottwo{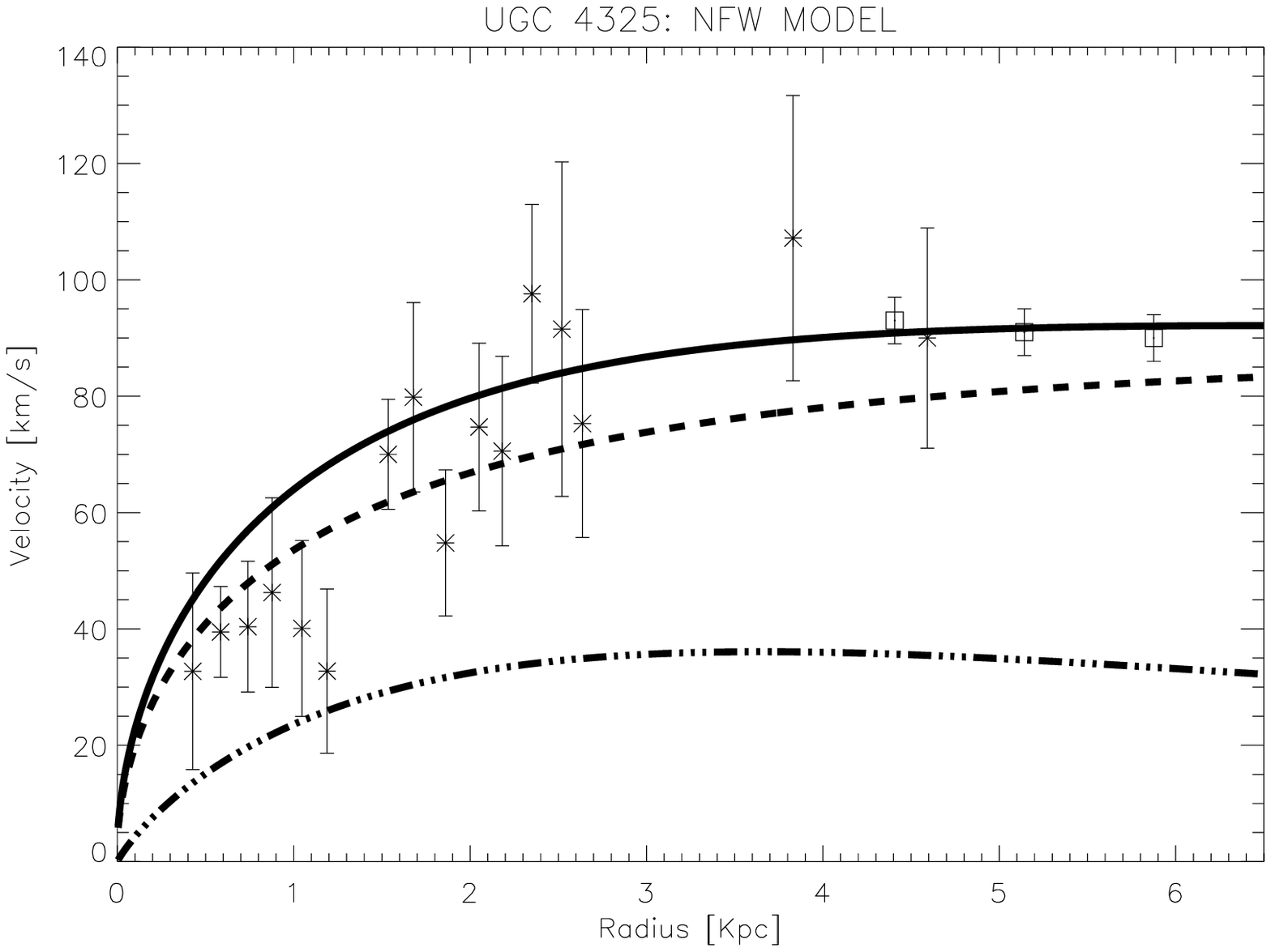}{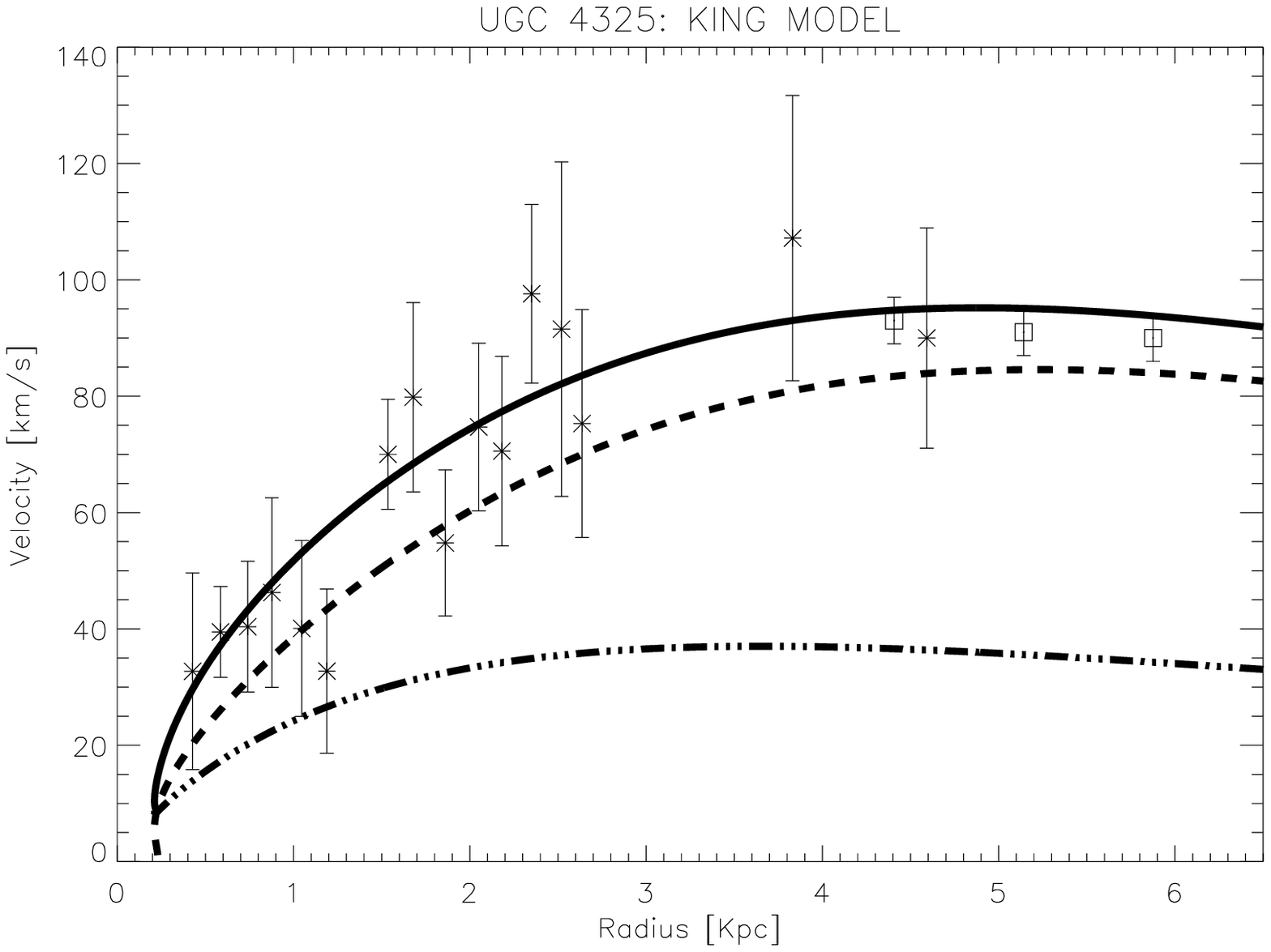}
\end{figure}

In Figure~\ref{f18.psf19.psf20.psf21.psf22.psf23.psf24.psf25.psf26.psf27.psf28.psf29.ps} the analysis for the six galaxies of our sample is shown. In the left-hand panels the NFW is assumed as the representative mass distribution of the primordial halo (dotted lines), while in the right-hand panel of each galaxy a King model is adopted (dotted lines). The disk is represented by the dot-dashed lines and the final rotation curve, corrected for the halo adiabatic contraction, is the solid line. Asterisks are our high-spatial resolution H$\alpha$ rotation curves combined in the outer parts with the HI data (squares).

It is clear that all the observed rotation curves are reproduced best when the dark matter in the primeval halo is distributed as a King profile. The NFW profile is always inconsistent with the inner parts of the observed rotation curves, as it predicts velocities which rise too fast.

\subsection{Scaling relations}

\citet{fi01} analyzed a sample of dark matter dominated galaxies (dwarfs and LSBs) with accurate HI rotation curves taken from the literature. All these galaxies showed slowly rising rotation curves and evidence for soft cores. The authors correlated the halo scales from galaxies to galaxy clusters, including in the analysis the halo scales inferred from CL0024+1654 and other clusters with a suspect evidence of flat mass distribution at the center. Surprisingly they found that the halo central density is independent on the halo mass, ranging from dwarf galaxies to galaxy clusters and the core radius scales with the mass. However, due to the low spatial resolution of the radio observations, the HI data are potentially affected by beam smearing, which could mask the real mass distribution in the halo inner regions.

With the optical rotation curves for dark matter dominated galaxies, we correlate the halo scales from dwarf galaxies to galaxy clusters. Since the halos of these galaxies seem to be core dominated and well fitted by a King model with form parameter $f_{p}=8$, we use the estimated values for the scales $\rho_{o}$ (central density) and $r_{o}$ (core radius) of the King profile derived in the previous section and corrected for halo adiabatic contraction in order to test the robustness of the scaling relations found by \citet{fi01}.

\begin{figure}
\caption{The halo central density and the core radius are showed as a function of the maximum circular velocity for LSB and dwarf galaxies inferred by HI rotation curves taken from the literature (dark squares); the white circles are the optical high resolution rotation curves by Swaters, Madore $\&$ Trewella (2000) but corrected for adiabatic contraction. Dark circles are
our sample of high spatial resolution rotation curves. The two independents points 
on the right of the plot are the galaxy clusters: CL0024+1654
and A1795. The dashed line corresponds to the prediction in \citet{sell00}. \label{f30.ps}}
\vskip 0.5truecm
\plotone{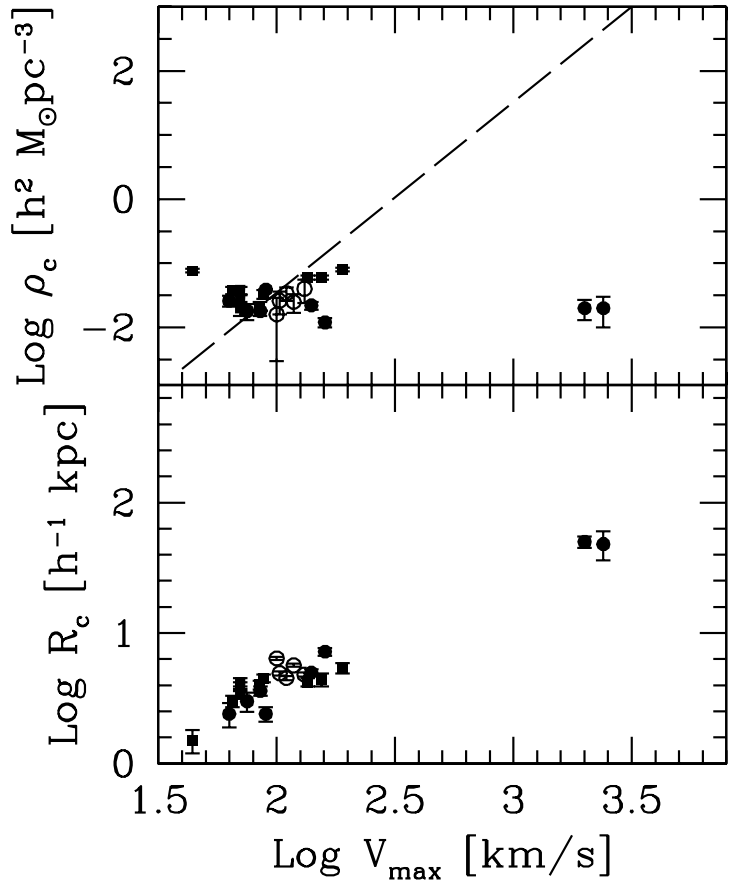}
\end{figure}

In Figure~\ref{f30.ps} the halo central density and the core radius are shown as functions of the maximum circular velocity for LSB and dwarf galaxies inferred by HI rotation curves taken from the literature (dark squares). The white circles are the optical high resolution rotation curves by \citet{SwMaTr00} but corrected for adiabatic contraction. Dark circles are our sample of high spatial resolution rotation curves. The two independent points on the right of the plot are the galaxy clusters: CL0024+1654 and A1795.

The scale invariance (within a factor of 2) of the halo central density is preserved and the core radius scales proportional to the halo mass. The optical rotation curves confirm the scale properties estimated in \citet{fi01}. Further observations are needed especially on galaxy cluster scales, for which there are still many uncertainties; furthermore, soft cores are not found at the center of every galaxy cluster (David et al. 2000) or they are very small (Arabadjis, Bautz \& Garmire 2001). Remarkably, if the core existence is confirmed on galaxy cluster scales the CDM models are unable to predict the scale invariance of the halo central density.

If future observations will confirm this scale invariance of the halo central density, it has interesting implications for the nature of the dark matter. If the dark matter is assumed to be warm, the maximum space phase density, defined as $f_{max}=\rho_{o}/\sigma^{3}$ with $\rho_{o}$ the halo central density and $\sigma$ the halo velocity dispersion ($\sigma \propto v_{max}$), has a finite value. As a consequence of the Liouville's theorem $f_{max}$ is preserved, implying an increase of the halo central density for more massive halos: $\rho_{o} \sim v_{max}^{3}$ (dashed line in the top panel of Figure~\ref{f30.ps}). This is inconsistent with the constant central density shown here. Indeed, if we accept the scale invariance of the halo central density on the halo mass ($\rho_{o} \approx$ const), this rules out the fermionic warm particles as candidates for the  dark matter. On the other hand, if the dark matter is assumed to be weakly self-interacting, the scaling relations shown in this paper can be reproduced if the cross section is assumed to be inversely proportional to the halo dispersion velocity (D'Onghia, Firmani \& Chincarini 2002). However, the self-interacting dark matter suffers from other conflicts that make this scenario unlikely to solve all the problems of the CDM models (e.g. Gnedin \& Ostriker 2001).

\section{CONCLUSIONS}

From the analysis of high spatial resolution rotation curves obtained with
H$\alpha$ long slit spectroscopy of four late-type dwarf galaxies and two LSB 
galaxies we reach the following conclusions:
\begin{itemize}
\item By comparing the H$\alpha$ and the HI rotation curves in the inner regions, we find a good agreement for all the six galaxies except for LSB F571-8. In other words, the HI data available in the literature for these galaxies are not substantially affected by the modest spatial resolution of the radio observations.
The galaxy F571-8 is seen edge-on so that different optical depth and projection effects may easily alter the observed rotation curve.
\item In the minimum disk hypothesis, in which we really neglect the disk contribution to the total gravitational potential, the observed rotation curves match better to the soft core density profile (Burkert) than to cuspy profiles (NFW and Moore). The profiles predicted by CDM N-body simulations are severely inconsistent in the inner regions, predicting rotation curves rising too fast within a few kpc.
\item By forcing the fit in the minimum disk hypothesis with the NFW profile we
were able to fit the observations reasonably well; however, the best fit has maximum velocities that are up to 30$\%$ higher than the velocity at the largest observed radius.
\item Accounting for the presence of baryonic matter in the nucleus, we find that the King profile (characterized by a soft density
core) fits well to the primeval dark matter halo density profile derived from our rotation curves. This is not the case when the
primeval dark matter halo density profile is modeled with a NFW profile. If a Moore profile is assumed for the initial halo, the discrepancy is exacerbated.
\end{itemize}

Finally, the optical rotation curves confirm for LSB and late type dwarf galaxies halo central density close to the value of 0.05 h$^{2}$ $M_{\odot}$pc$^{-3}$ as estimated previously on the basis of the HI data.

\section{ACKNOWLEDGMENTS}

We are grateful to the referee, Stacy McGaugh, for very helpful and detailed comments and for very interesting discussions. DM wishes to thank Marcella Longhetti for her contribution in the reduction and interpretation of the data; Paolo Saracco for constructive suggestions; and all the staff of the TNG for their continuing support during the observations and for their great hospitality. ED thanks Andi Burkert for stimulating discussions on the topic and for an early reading of the draft. DM and ED thank Fondazione Cariplo for financial support.

\clearpage

\begin{deluxetable}{ccccccccc}
\tabletypesize{\scriptsize}
\tablecaption{General Properties of the Galaxies \label{tbl-1}}
\tablewidth{0pt}
\tablehead{
\colhead{Name} & \colhead{D\tablenotemark{a}} & \colhead{h\tablenotemark{b}} & \colhead{i\tablenotemark{c}} &
\colhead{P.A.\tablenotemark{c,d}} & \colhead{$v_{sys}$\tablenotemark{c}} & \colhead{$\mathrm{\mu}_{o}$\tablenotemark{b}} & \colhead{Morph. Type} & \colhead{Refs\tablenotemark{e}}\\
 & (Mpc) & (kpc) & ($^{\circ}$) & ($^{\circ}$) & (km s$^{-1}$) & (m arcsec$^{-2}$)& & \\
(1) & (2) & (3) & (4) & (5) & (6) & (7) & (8) & (9)\\
}
\startdata
UGC4325   &10.1& 1.6 (R) & 41 & 41  & 523  & 21.6 (R) & SA(s)m  & (1)(4)\\
UGC4499   &13.0& 1.5 (R) & 50 & 140 & 691  & 21.5 (R) & SABdm   & (1)(4)\\
UGC7603   &6.8 & 0.9 (R) & 78 & 197 & 644  & 20.8 (R) & SB(s)dSp& (1)(4)\\
UGC11861  &25.1& 6.1 (R) & 50 & 28  & 1481 & 21.4 (R) & SABdm   & (1)(4)\\
LSB F571-8&48.0& 5.2 (B) & 90 & 165 & 3754 & 23.9 (B) & Sc      & (2)(3)\\
LSB F583-1&32.0& 1.6 (B) & 63 & 175 & 2264 & 24.1 (B) & Sm-Irr  & (2)(3)\\
\enddata

\tablenotetext{a}{Using $H_{0}$=75 km s$^{-1}$ Mpc$^{-1}$}
\tablenotetext{b}{(B)=B filter photometry~~~(R)=R filter photometry}
\tablenotetext{c}{From HI observations in literature}
\tablenotetext{d}{Measured positive from North to East}
\tablenotetext{e}{(1)Swaters (1999) (2)de Blok et al. (1997) (3)de Blok et al. (2001) (4)van den Bosch et al. (2001)}
\end{deluxetable}

\clearpage

\begin{deluxetable}{ll}
\tabletypesize{\scriptsize}
\tablecaption{Instrumental parameters \label{tbl-2}}
\tablewidth{0pt}
\tablehead{
\colhead{Parameter} & \colhead{Value}
}
\startdata
Spectrograph & {\it d.o.lo.res.} \\
Grism & HR-r\\
Spectral coverage & 6200-7800 \AA \\
Slit & 1$^{\prime \prime}$-1.5$^{\prime \prime}$ $\times$9.4$^{\prime}$ \\
Spatial Scale & 0.275$^{\prime \prime}$ pixel$^{-1}$~~~1 pixel=15$\mu$m \\
Dispersion & 0.8 \AA~pixel$^{-1}$ \\
Resolution (FWHM) & $\sim$3 \AA \tablenotemark{a} \\
Detector & Loral CCD \tablenotemark{b} \\
CCD dimensions & 2048$\times$2048 pixel$^{2}$ \\
Field of view & 9.4$^{\prime}$$\times$9.4$^{\prime}$ \\
Readout noise & $\sim$7 e$^{-}$ r.m.s. \\
Conversion Factor & $\sim$1 e$^{-}$/ADU \\
CCD QE peak & 95$\%$ at 6000 \AA \\
\\
\enddata

\tablenotetext{a}{For a 1 arcsec slit}
\tablenotetext{b}{Thinned and back-illuminated}
\end{deluxetable}

\clearpage

\clearpage

\clearpage

\begin{figure}
\vskip -0.5truecm
\plottwo{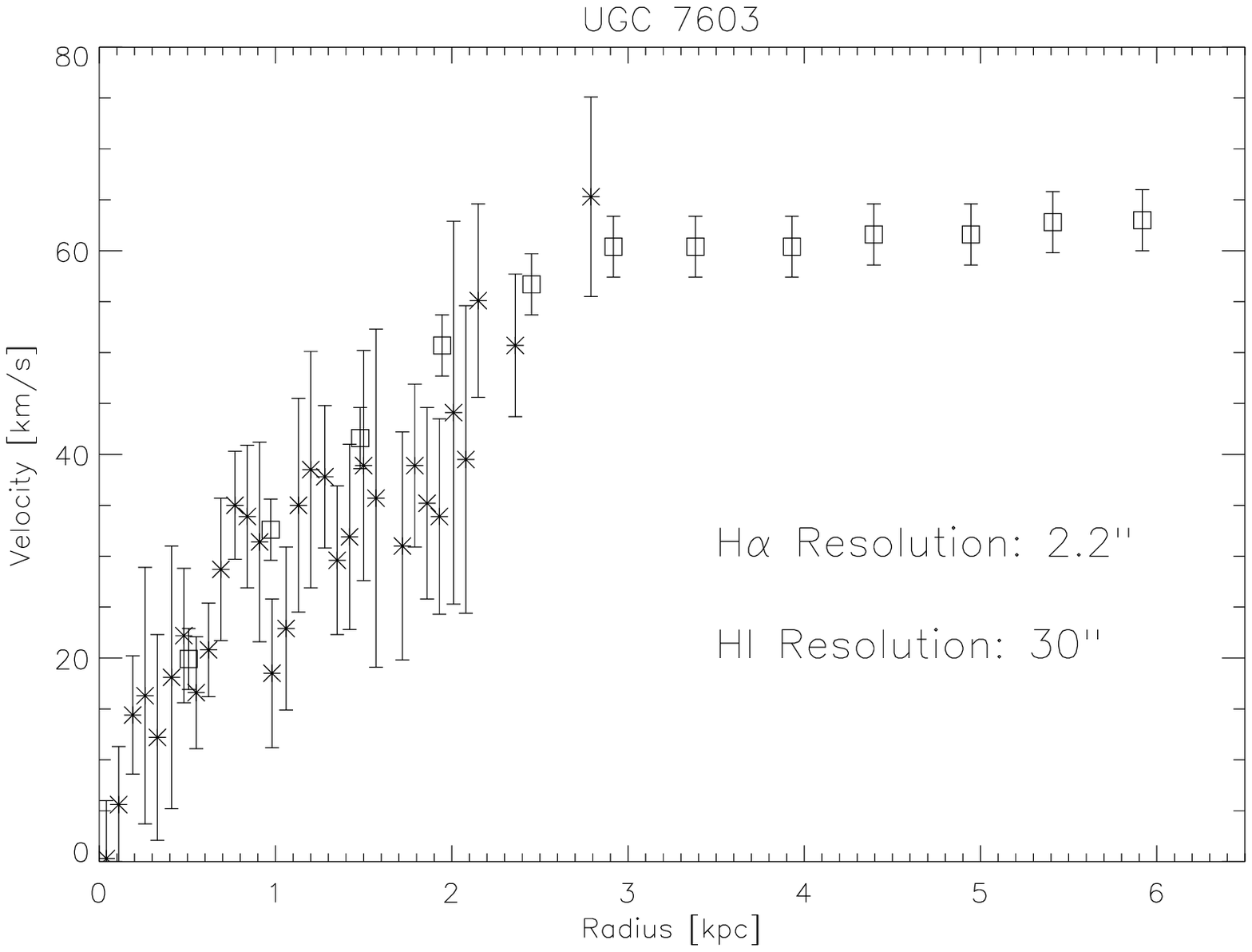}{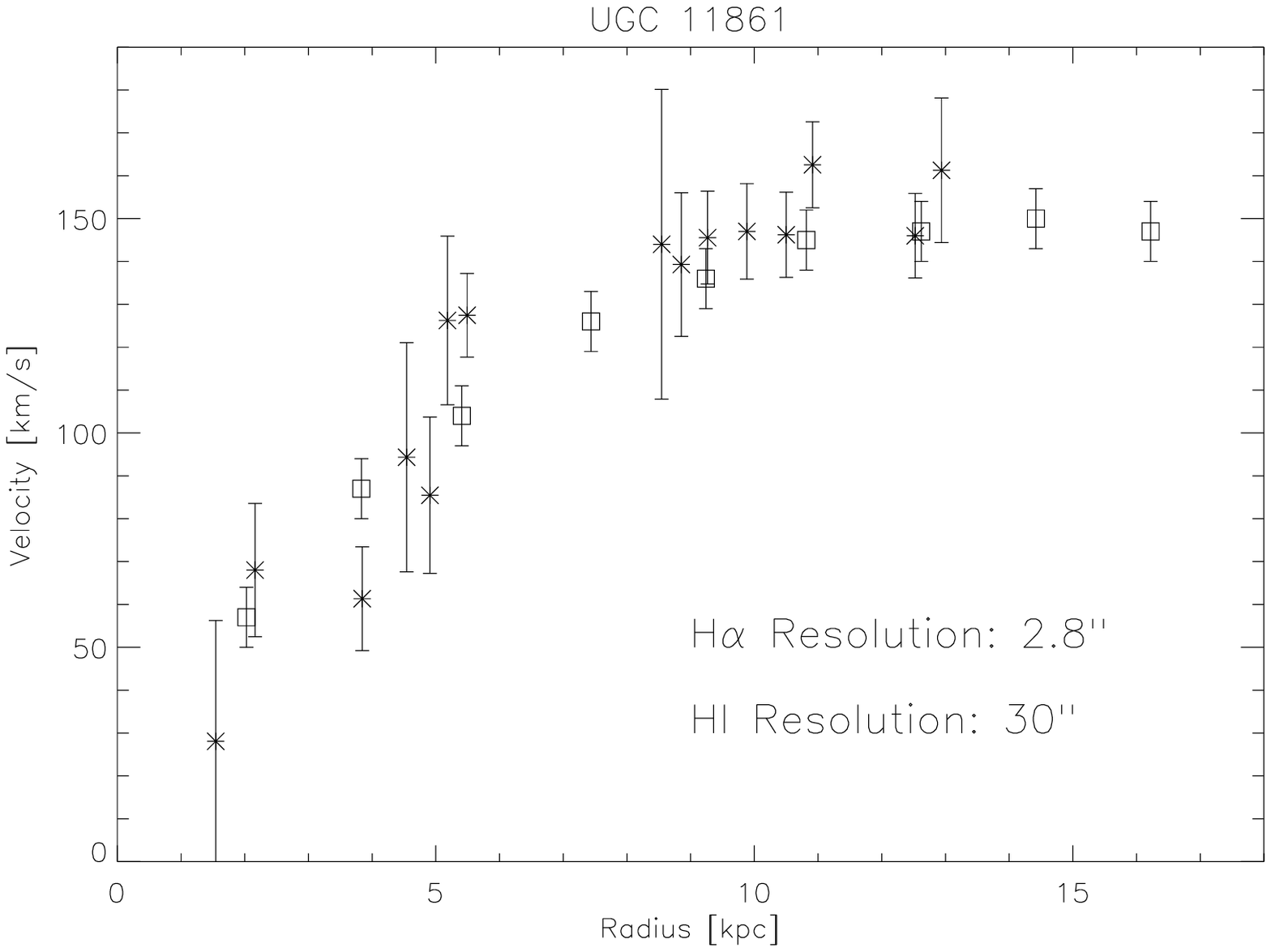}
\end{figure}
\begin{figure}
\vskip -0.5truecm
\plottwo{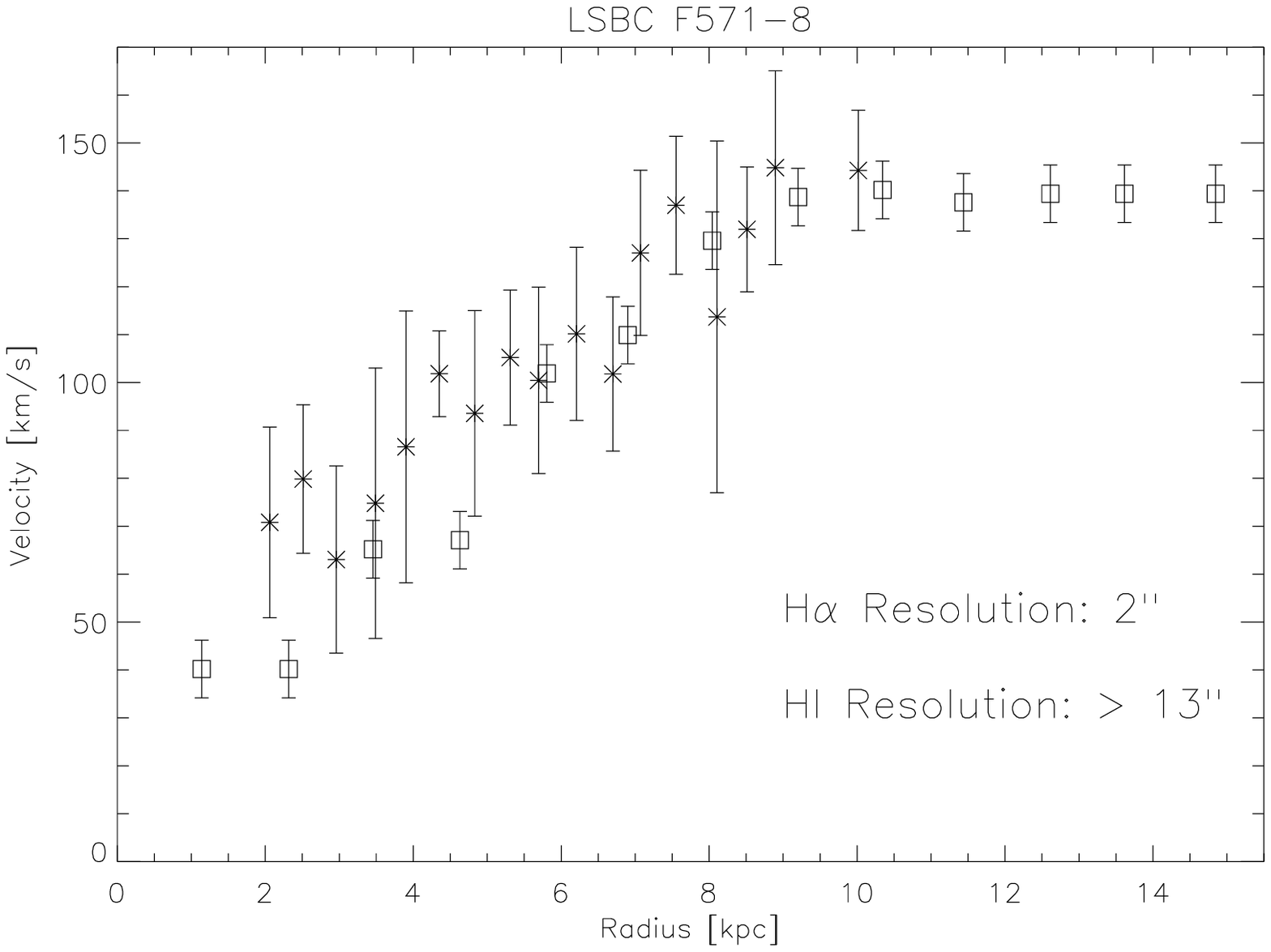}{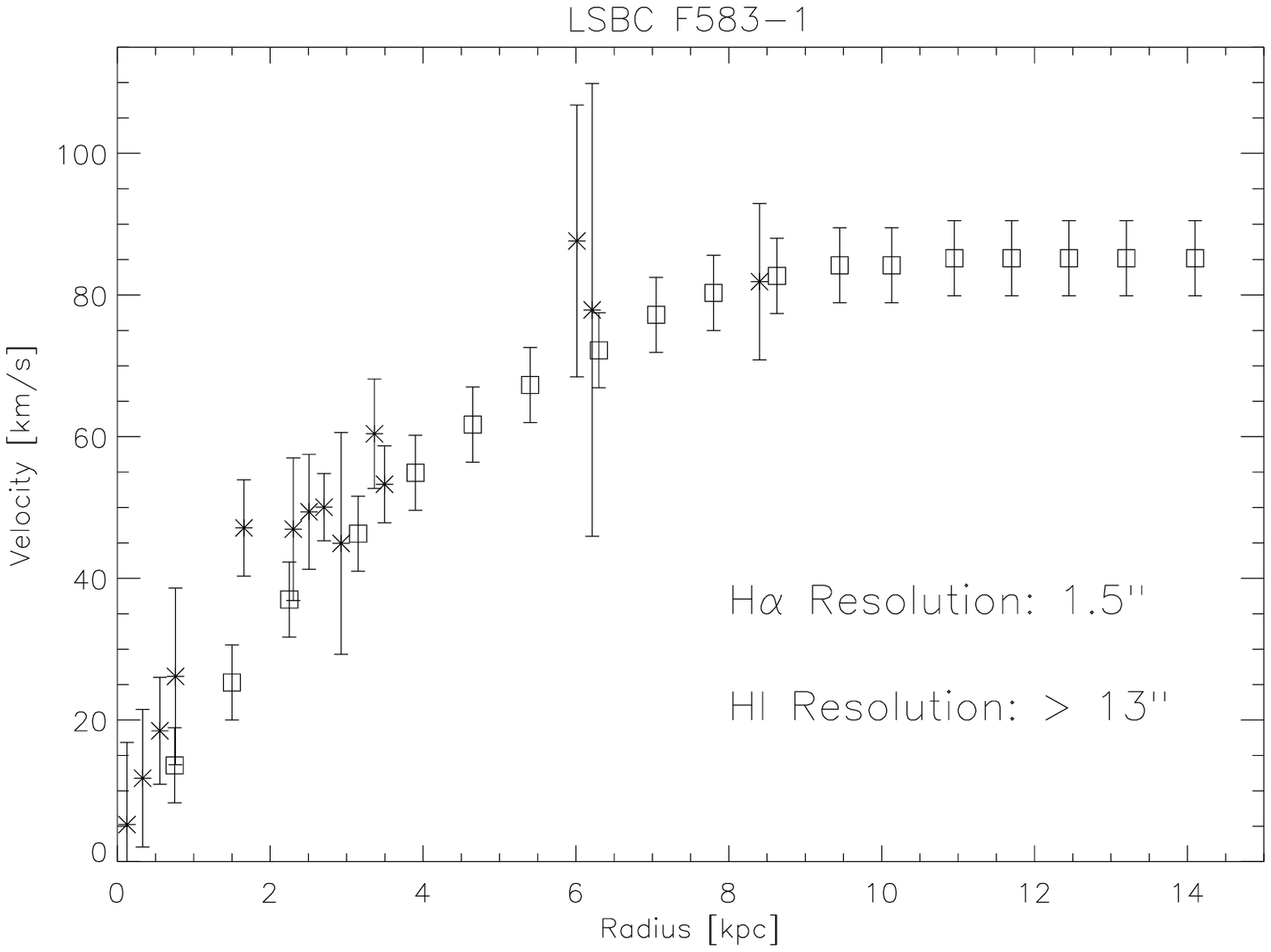}
\end{figure}

\clearpage

\begin{figure}
\vskip -0.5truecm
\plottwo{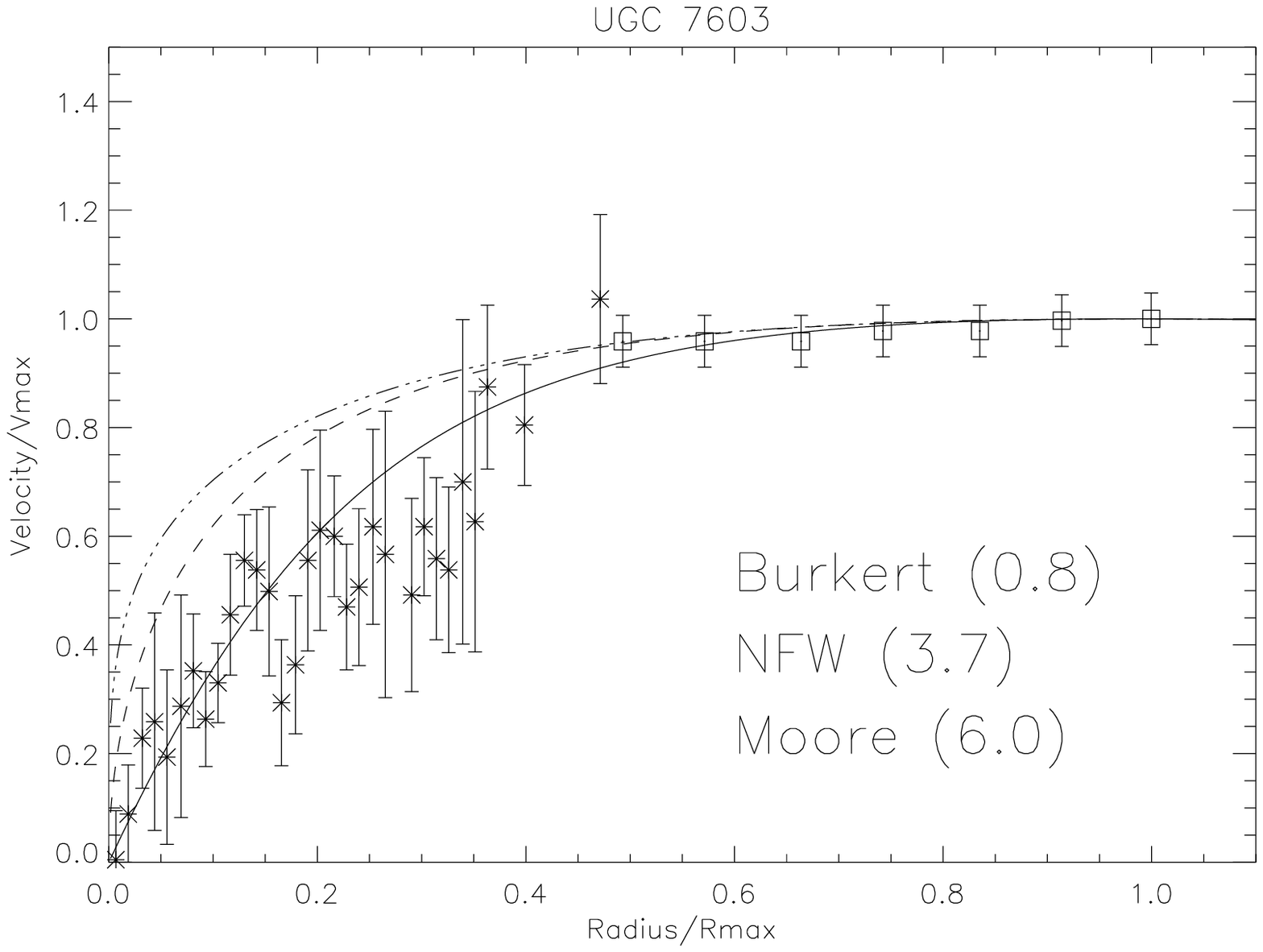}{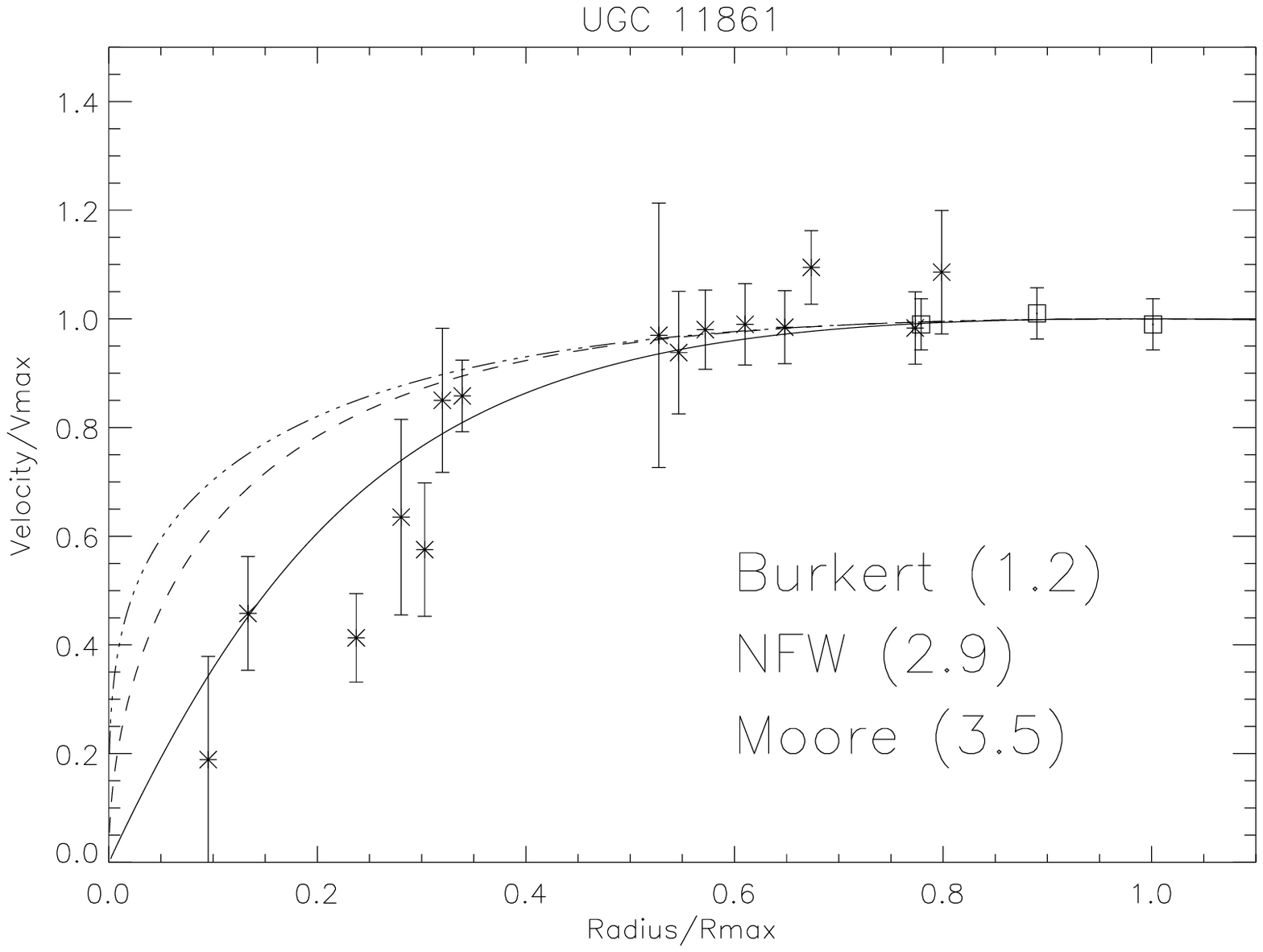}
\end{figure}
\begin{figure}
\vskip -0.5truecm
\plottwo{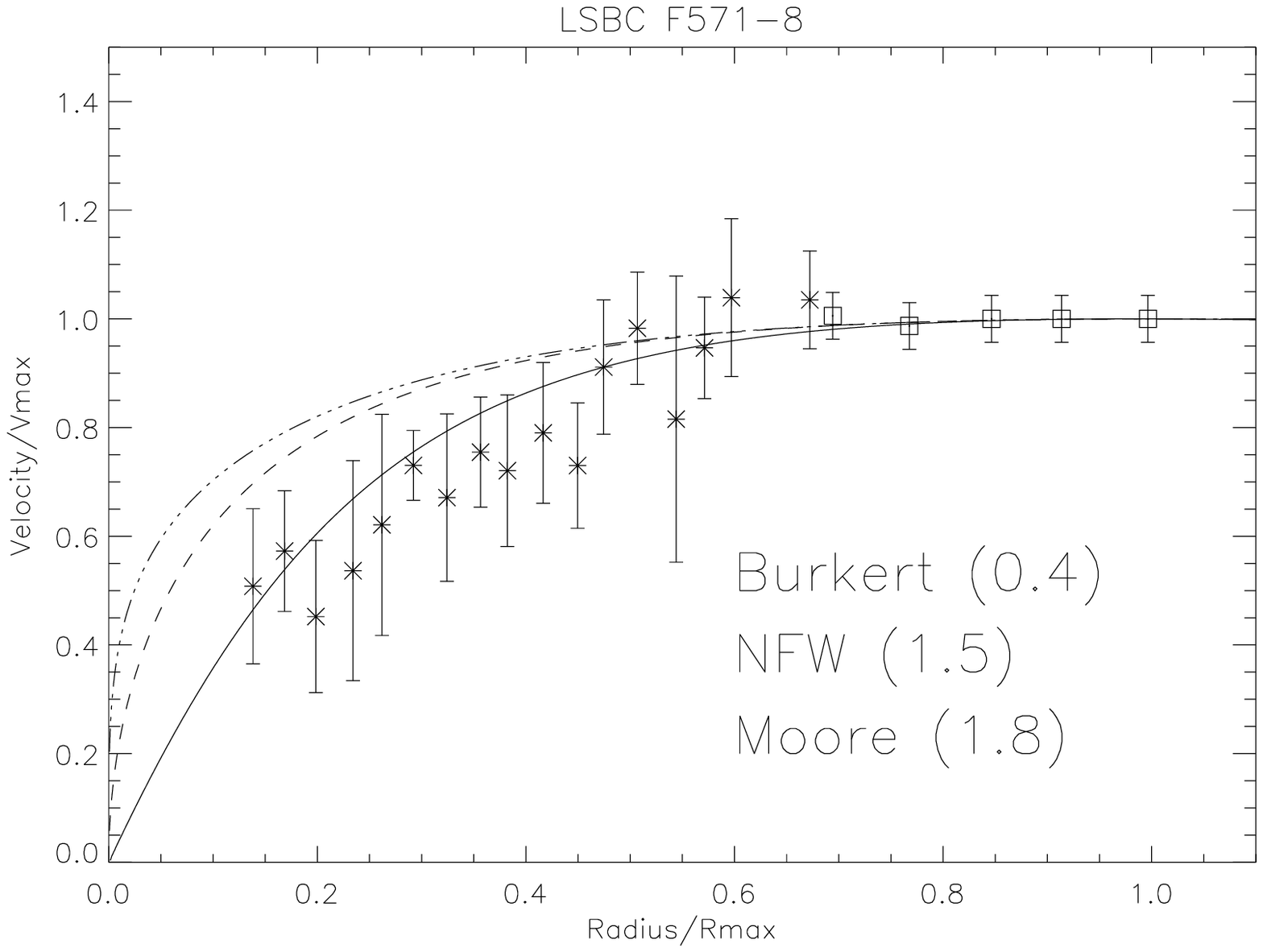}{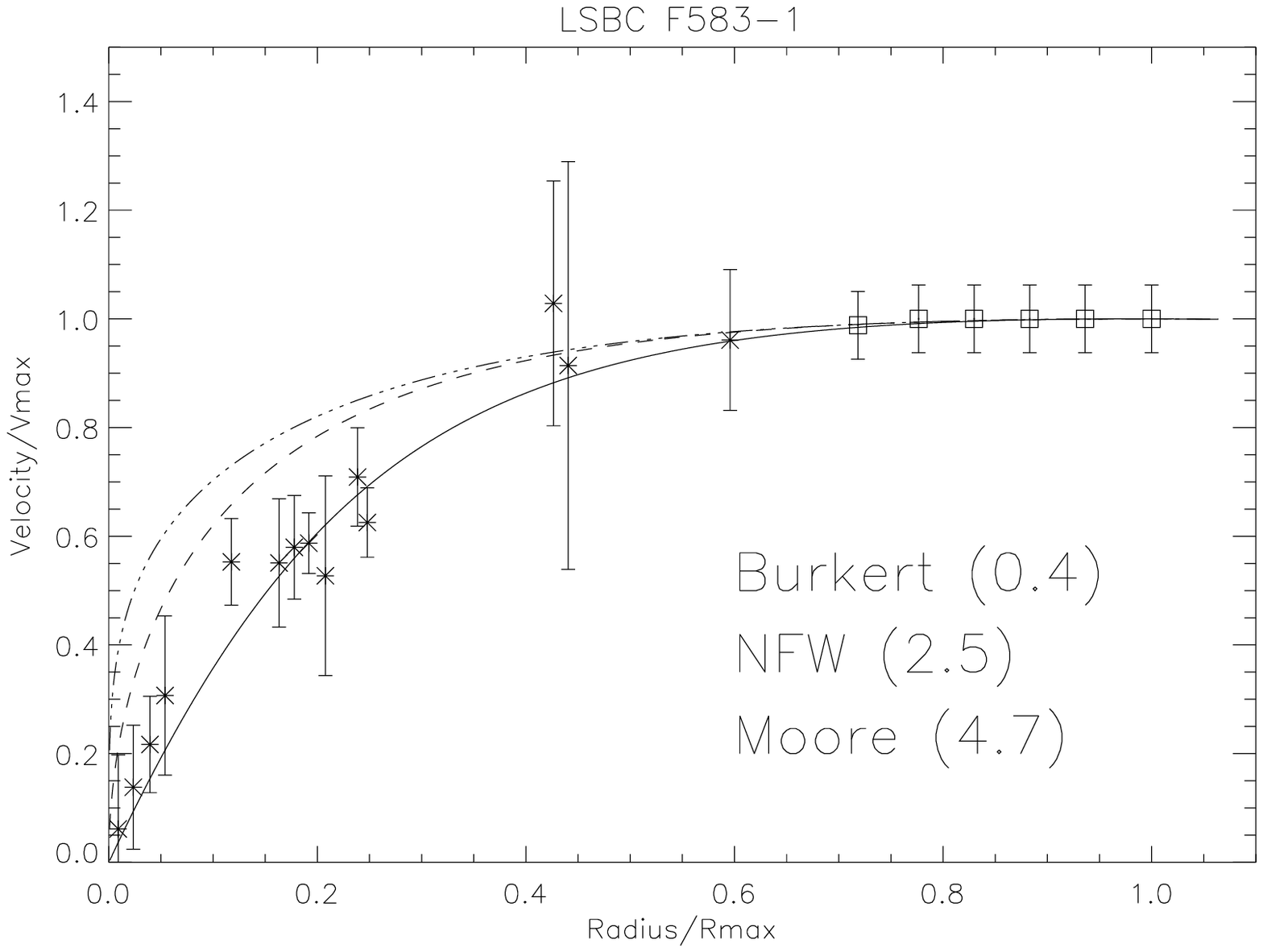}
\end{figure}

\clearpage

\begin{deluxetable}{cccccc}
\tabletypesize{\scriptsize}
\tablecaption{Best fit solution parameters for minimum disk model using the Burkert profile\label{tbl-3}}
\tablewidth{0pt}
\tablehead{
\colhead{Name} & \colhead{$\mathrm{\rho}_{o}$} & \colhead{$\Delta \mathrm{\rho}_{o}$\tablenotemark{a}} & \colhead{$\mathrm{r}_{o}$} & \colhead{$\Delta \mathrm{r}_{o}$\tablenotemark{a}} & \colhead{$\mathrm{\chi}_{\nu}^{2}$}\\
 & ($\mathrm{M}_{\odot}$~pc$^{-3}$) & ($\mathrm{M}_{\odot}$~pc$^{-3}$) & (kpc) & (kpc) &  \\
}
\startdata
UGC4325   &0.231& 0.044 & 1.77 & 0.18 & 0.6 \\
UGC4499   &0.090& 0.027 & 2.28 & 0.39 & 0.4 \\
UGC7603   &0.080& 0.016 & 2.09 & 0.68 & 0.7 \\
UGC11861  &0.060& 0.009 & 5.81 & 1.00 & 1.0 \\
LSB F571-8&0.065& 0.010 & 5.15 & 0.33 & 0.3 \\
LSB F583-1&0.035& 0.007 & 4.23 & 0.35 & 0.3 \\
\enddata

\tablenotetext{a}{Quoted errors are one-sigma errors on best fit solution parameters}
\end{deluxetable}

\begin{deluxetable}{crcrrcrc}
\tabletypesize{\scriptsize}
\tablecaption{Best fit solution parameters for minimum disk model using the NFW profile\label{tbl-4}}
\tablewidth{0pt}
\tablehead{
\colhead{Name} & \colhead{$c$} & \colhead{$\Delta c$\tablenotemark{a}} & \colhead{$\mathrm{v}_{200}$} & \colhead{$\Delta \mathrm{v}_{200}$\tablenotemark{a}} & \colhead{$\mathrm{\chi}_{\nu}^{2}$} & \colhead{$\Delta \mathrm{v}_{max}$\tablenotemark{b}} & \colhead{$\Delta \mathrm{v}_{last~HI}$}\tablenotemark{c} \\
 & & & (km s$^{-1}$) & (km s$^{-1}$) &  & (km s$^{-1}$) & (km s$^{-1}$)\\
}
\startdata
UGC4325   &13.2& 2.3 & 80  & 7  & 0.8 & 14 & 4 \\
UGC4499   & 8.8& 2.5 & 70  & 10 & 0.5 &  8 & 5 \\
UGC7603   & 5.8& 1.1 & 78  & 8  & 1.0 & 22 & 3 \\
UGC11861  & 7.4& 1.1 & 158 & 12 & 1.6 & 32 & 7 \\
LSB F571-8& 8.6& 1.3 & 136 & 10 & 0.5 & 20 & 6 \\
LSB F583-1& 5.4& 1.1 & 92  & 8  & 0.3 & 14 & 5 \\
\enddata

\tablenotetext{a}{Quoted errors are one-sigma errors on best fit solution parameters}
\tablenotetext{b}{This is the difference between the best fit solution maximum velocity and the observed maximum velocity of the rotation curve: the higher this value, the funnier the NFW best fit solution}
\tablenotetext{c}{This is the one-sigma velocity error of  the last observed point in the HI rotation curve}
\end{deluxetable}

\clearpage

\clearpage

\begin{deluxetable}{lccccccc}
\tabletypesize{\scriptsize}
\tablecaption{Parameters used in the modeling taking care of the adiabatic contraction and the stellar disk component\label{tbl-5}}
\tablewidth{0pt}
\tablehead{
\colhead{Name} & \colhead{h\tablenotemark{a}} & \colhead{$\mathrm{f}_{b}$} & \colhead{$\mathrm{\Upsilon}_{\star}$} & \colhead{$c$} & \colhead{$\mathrm{M}_{200}$} & \colhead{$\mathrm{\rho}_{o}$} & \colhead{$\mathrm{r}_{o}$} \\
 & (kpc) & (10$^{-2}$) & & & (10$^{10}$$\mathrm{M}_{\odot}$) & ($\mathrm{M}_{\odot}$~pc$^{-3}$) & (kpc) \\
}
\startdata
UGC4325   & 1.7 &1.6& 1.0 & 15.1$\pm$0.5 & 8.3$\pm$2.5  & 0.040$\pm$0.012 & 2.3$\pm$0.2 \\
UGC4499   & 2.1 &2.1& 1.0 & 16.1$\pm$0.4 & 4.2$\pm$1.0  & 0.018$\pm$0.003 & 2.9$\pm$0.2 \\
UGC7603   & 0.9 &0.8& 1.0 & 16.8$\pm$0.4 & 2.7$\pm$0.5  & 0.021$\pm$0.003 & 0.2$\pm$0.2 \\
UGC11861  & 6.1 &5.8& 1.0 & 13.3$\pm$0.3 & 29.1$\pm$6.3 & 0.010$\pm$0.002 & 7.5$\pm$0.7 \\
LSB F571-8& 5.2 &1.1& 1.4 & 12.9$\pm$0.3 & 38.7$\pm$6.8 & 0.024$\pm$0.002 & 5.0$\pm$0.2 \\
LSB F583-1& 1.6 &0.4& 1.4 & 15.1$\pm$0.3 & 7.9$\pm$2.1  & 0.013$\pm$0.002 & 4.4$\pm$0.2 \\
\enddata
\tablenotetext{a}{Stellar disk scale length; for UGC4499 we also used $\mathrm{r}_{H}$=0.3 kpc as the scale radius of the stellar bulge}
\end{deluxetable}

\clearpage

\begin{figure}
\vskip -0.5truecm
\plottwo{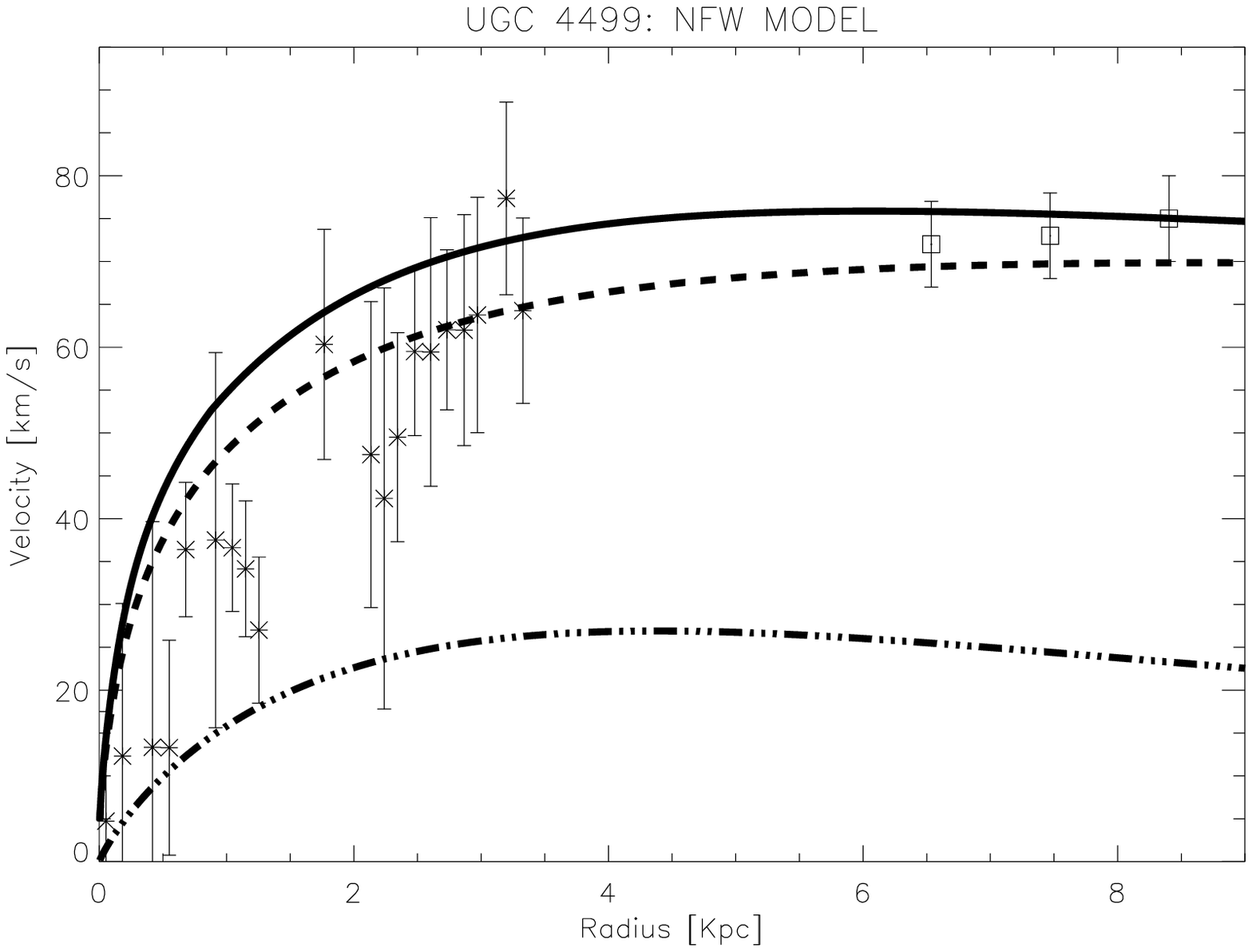}{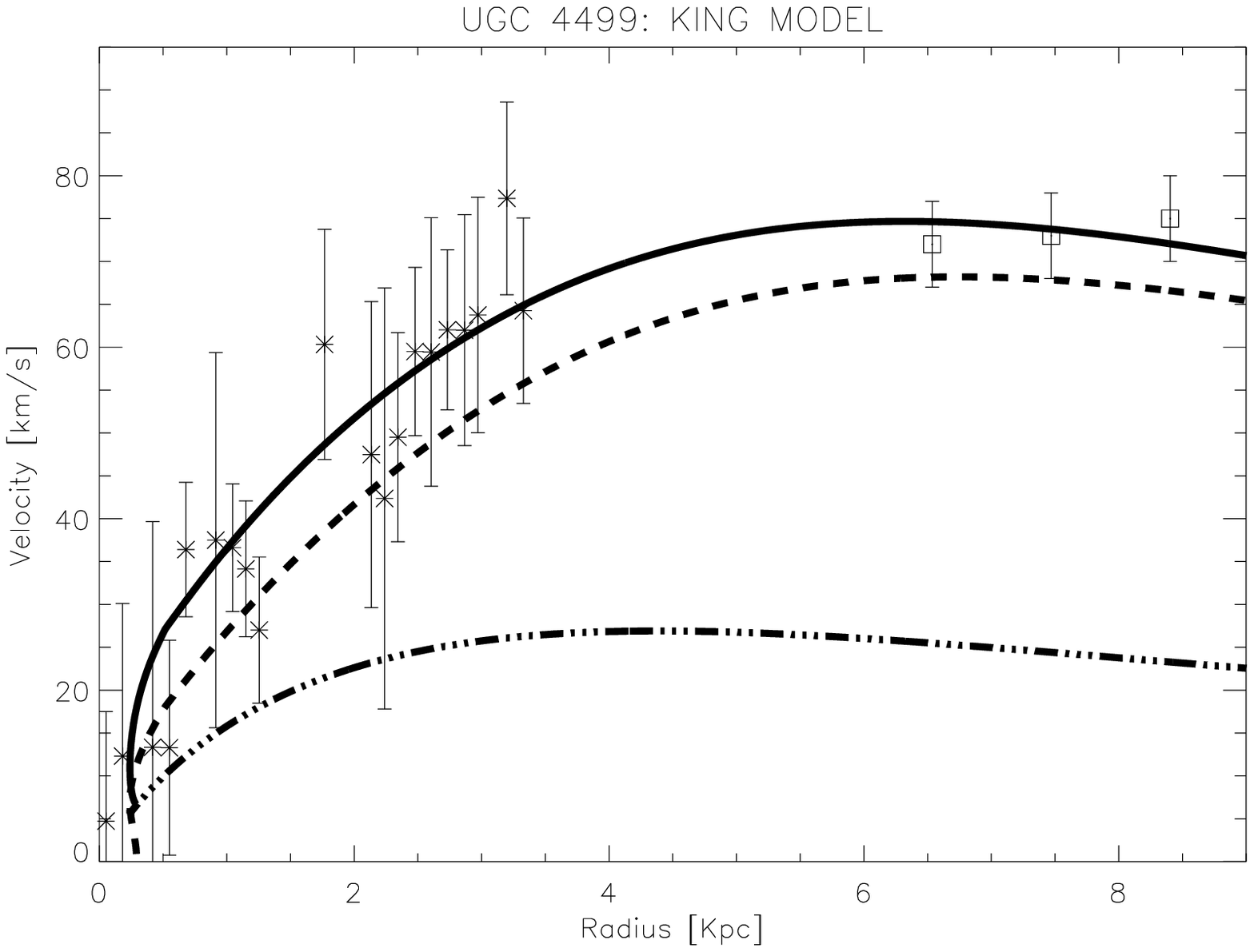}
\end{figure}
\begin{figure}
\vskip -0.5truecm
\plottwo{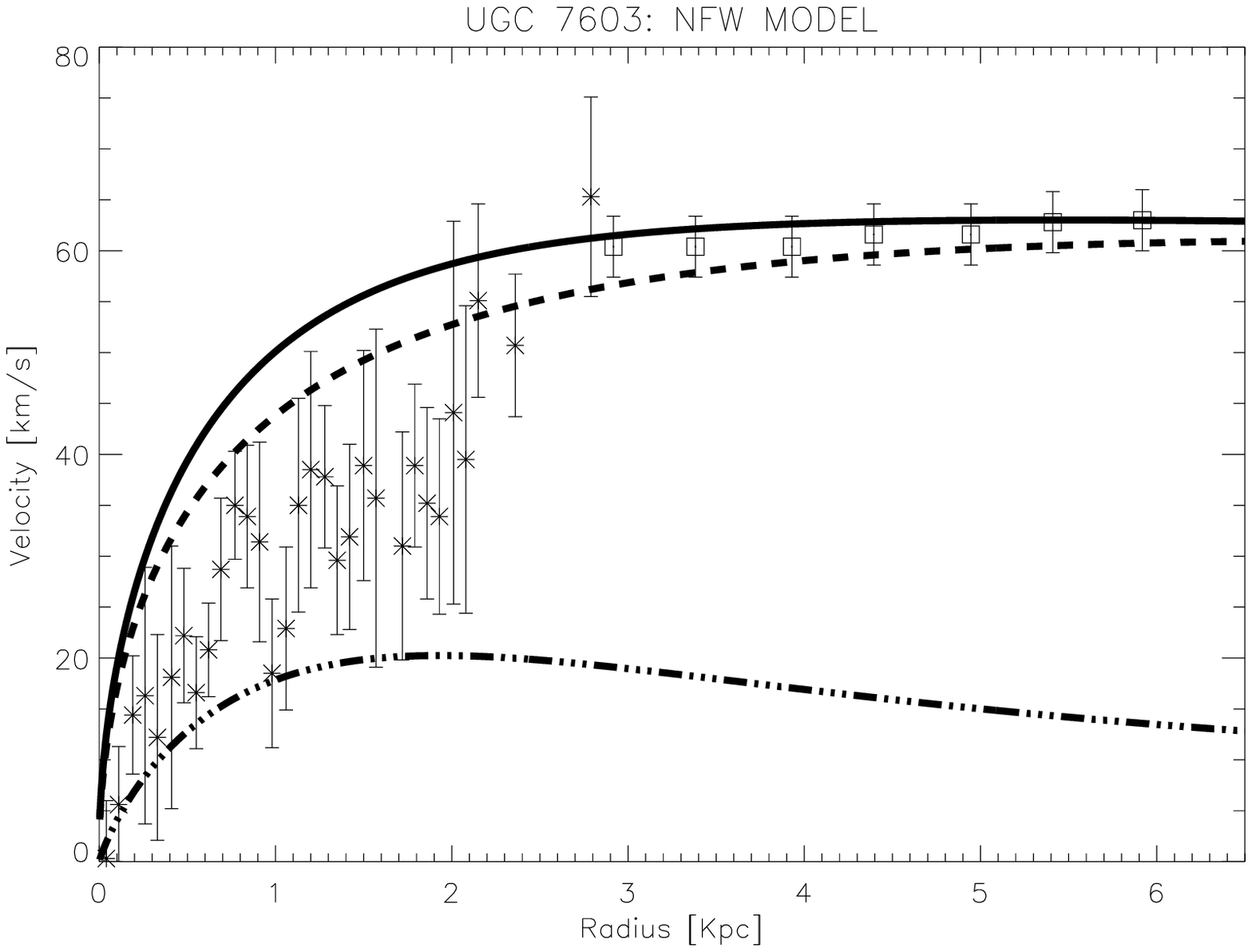}{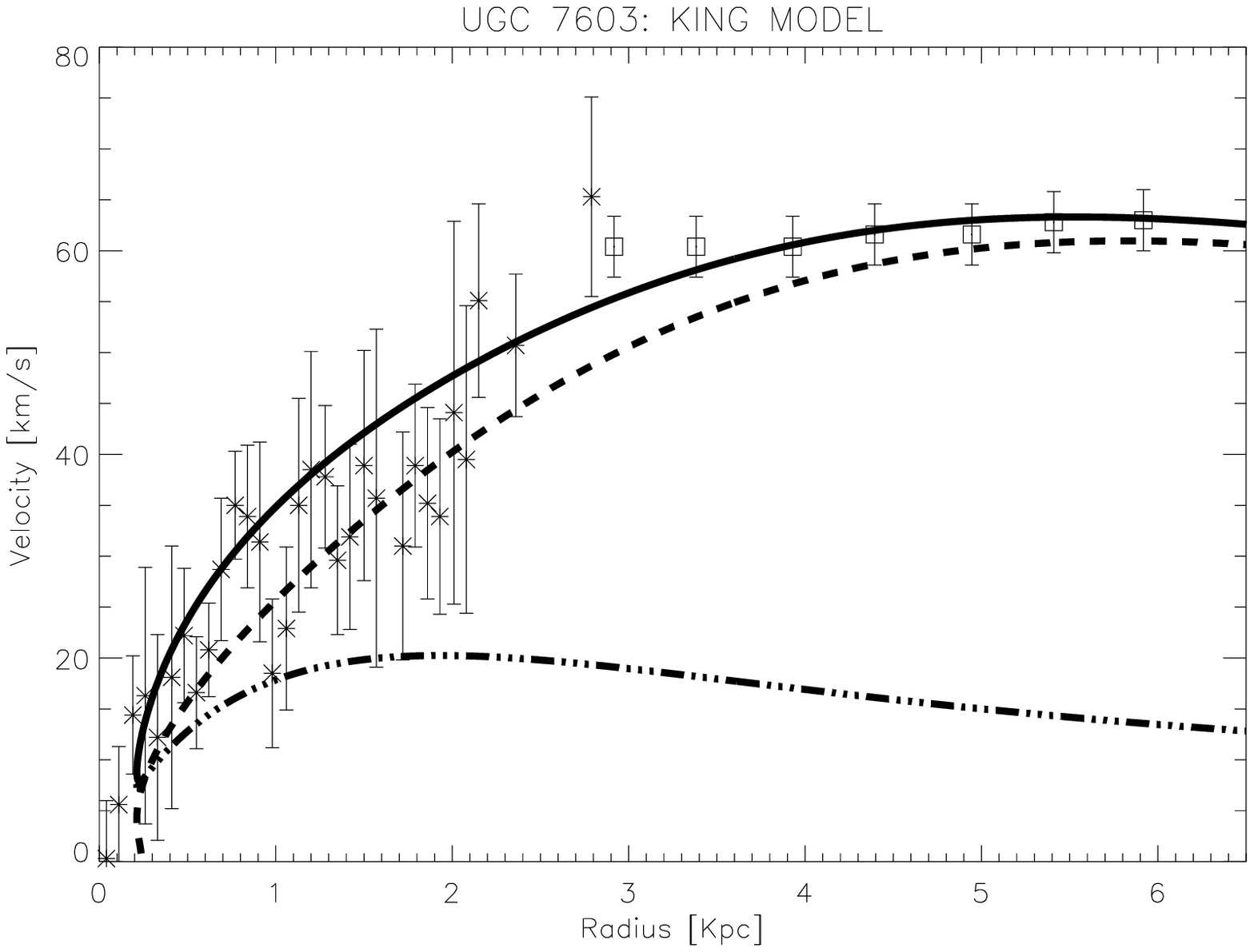}
\end{figure}
\begin{figure}
\vskip -0.5truecm
\plottwo{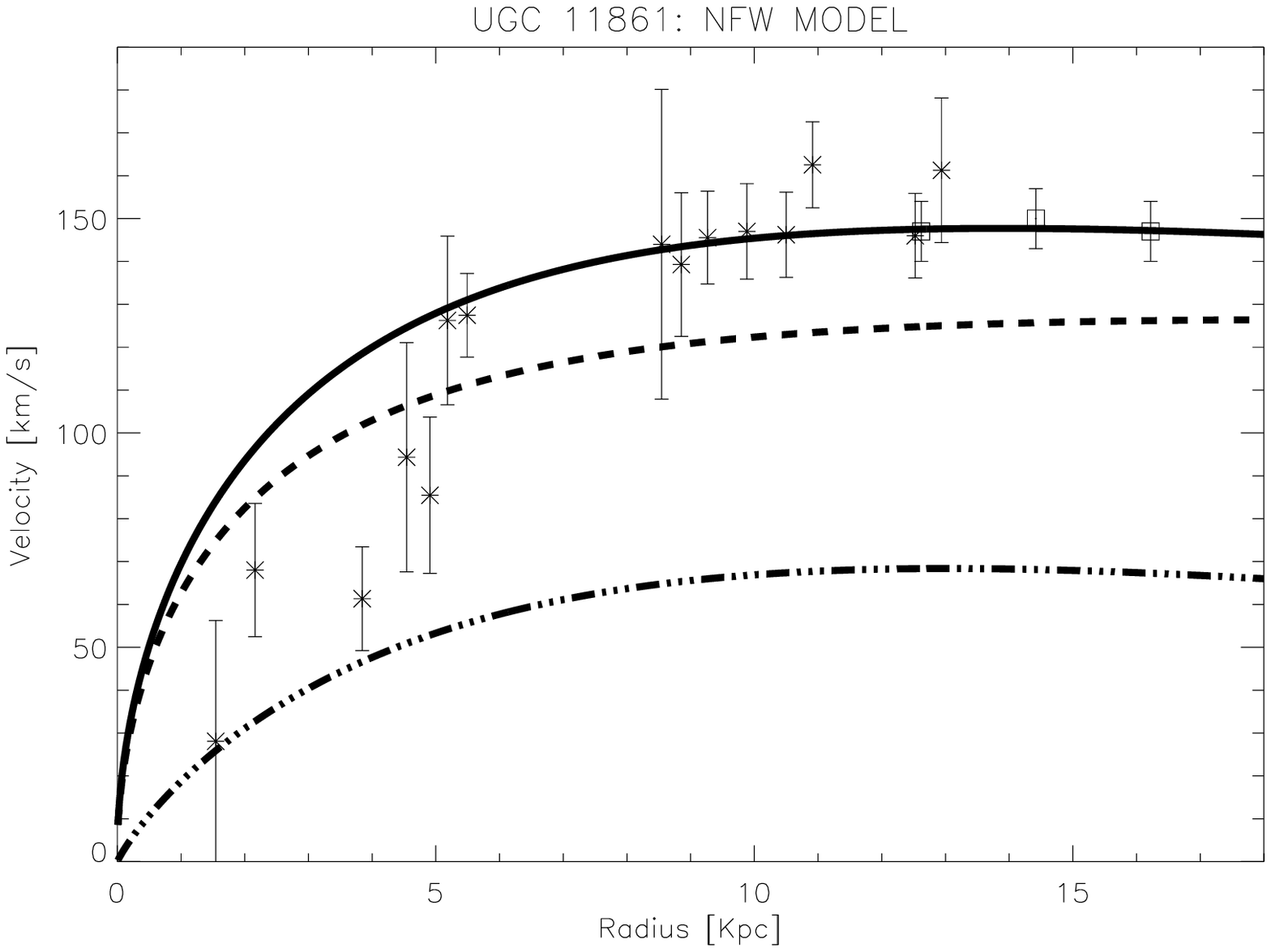}{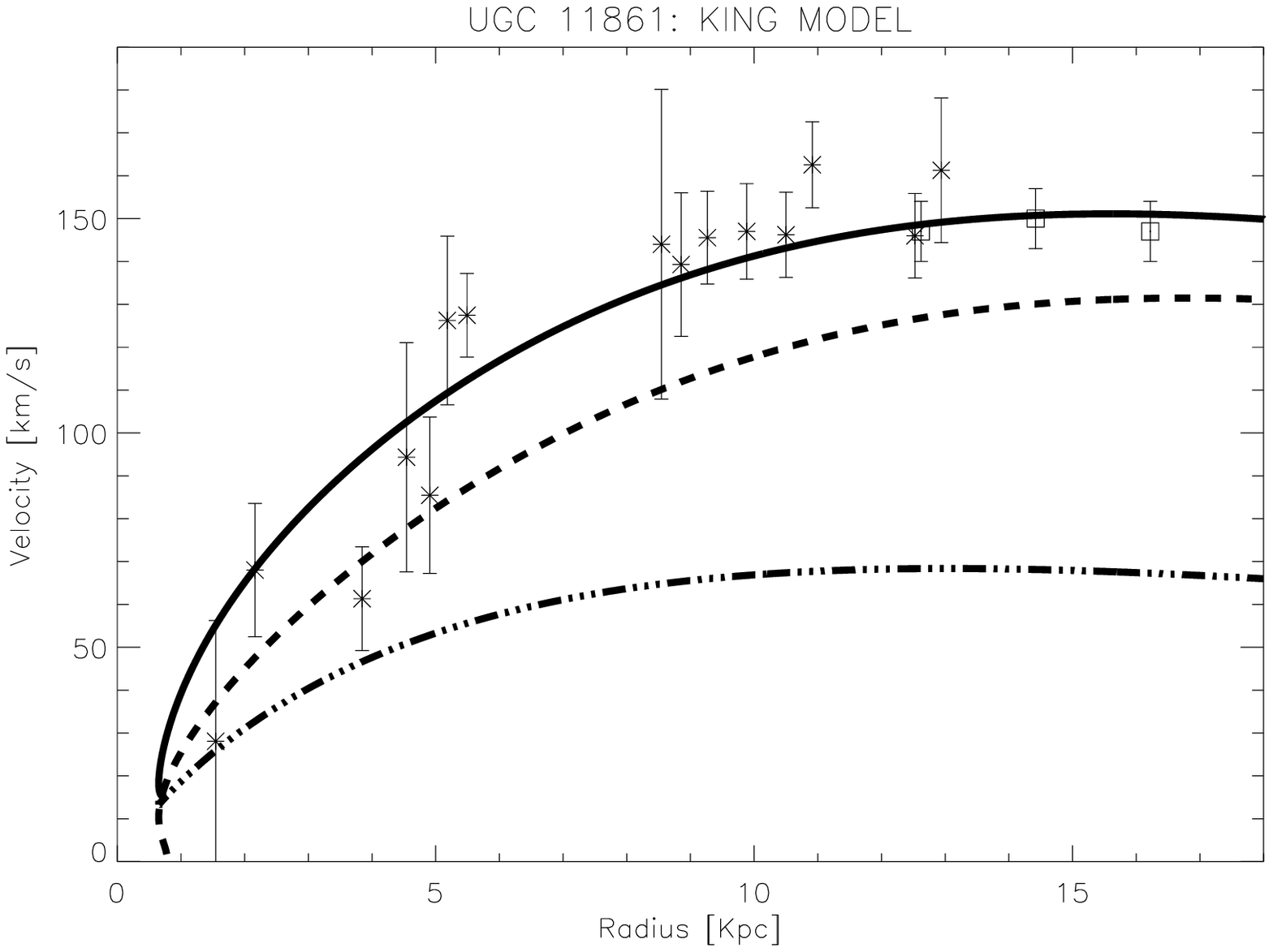}
\end{figure}
\begin{figure}
\vskip -0.5truecm
\plottwo{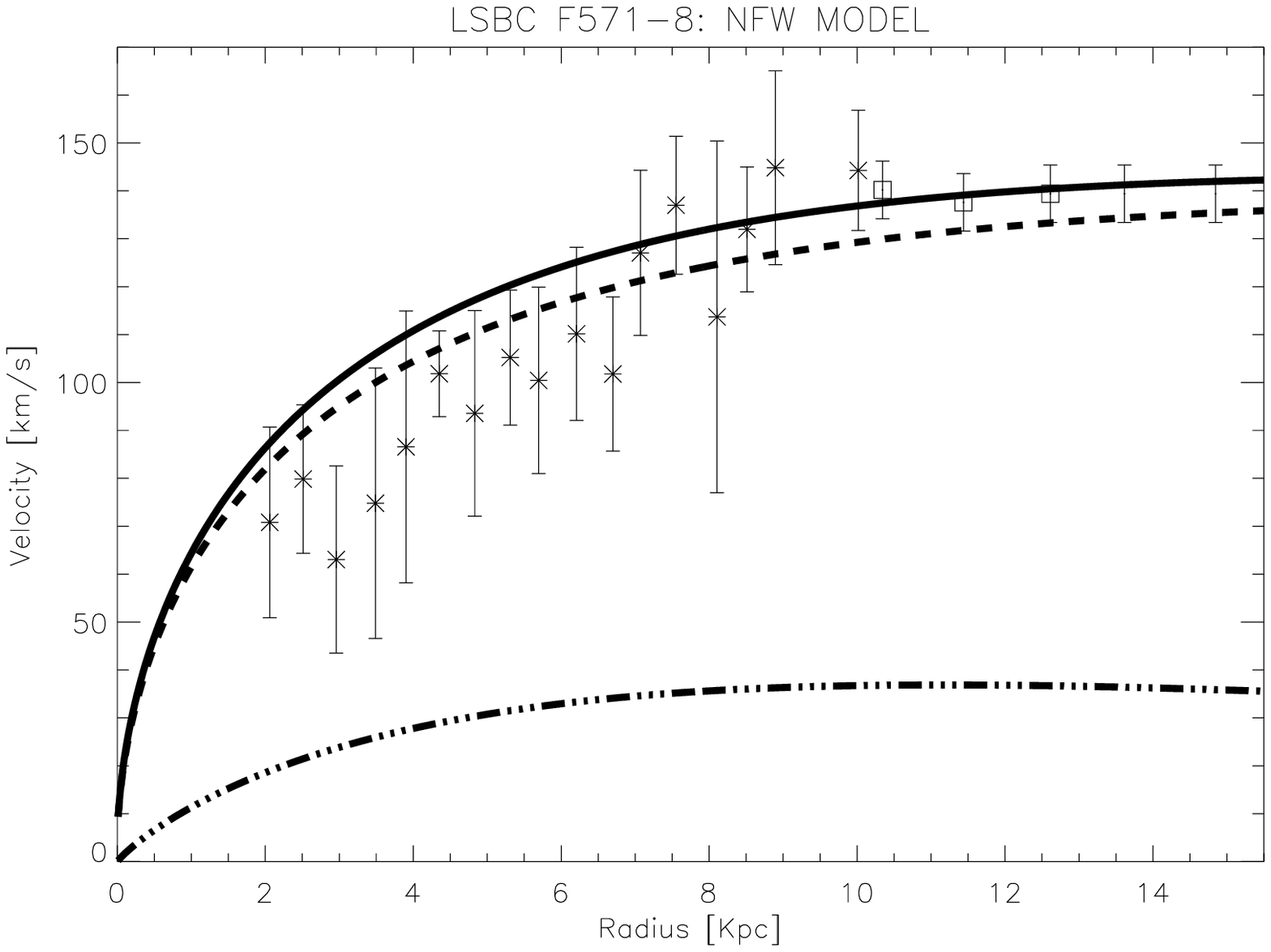}{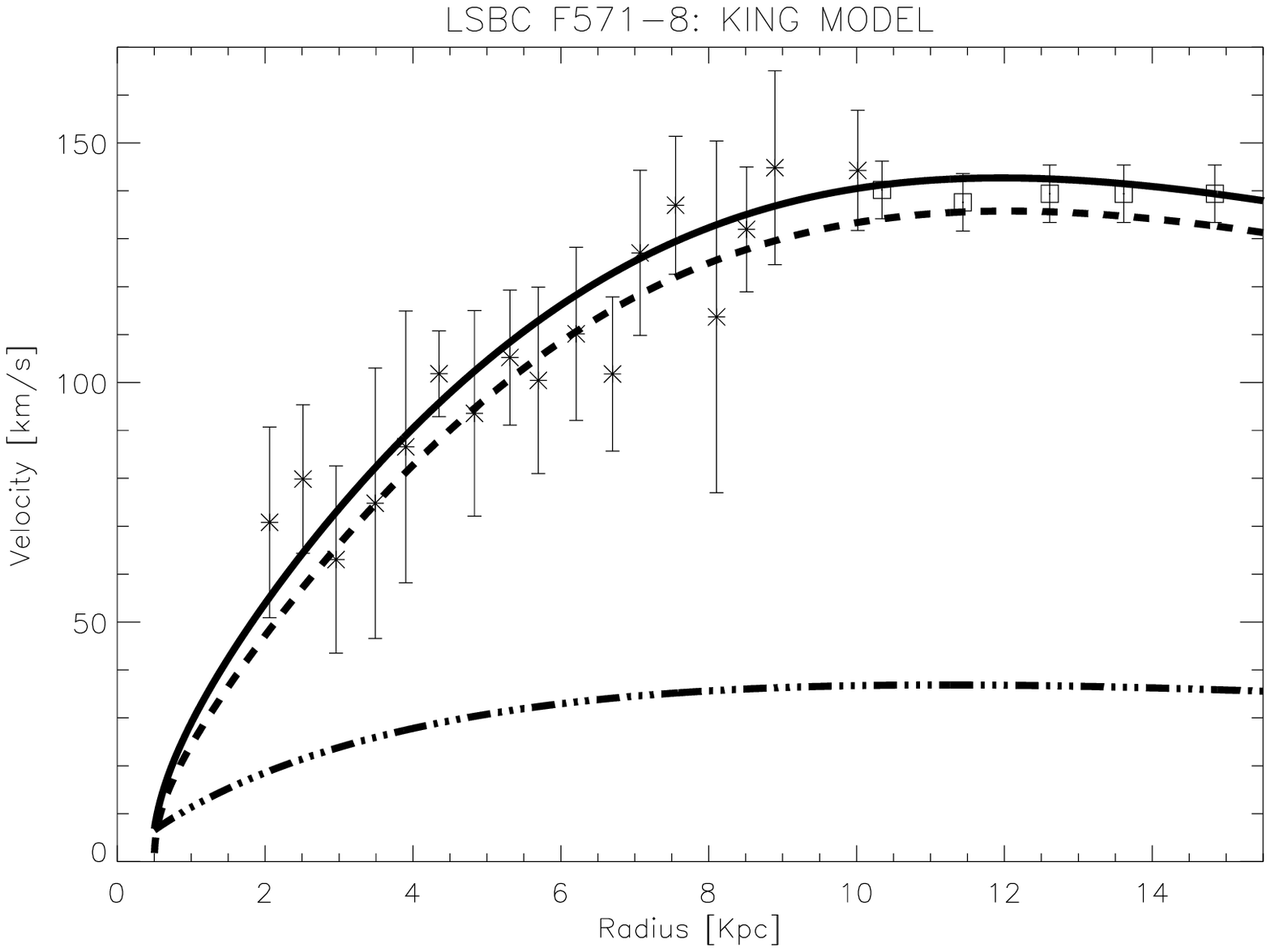}
\end{figure}
\begin{figure}
\vskip -0.5truecm
\plottwo{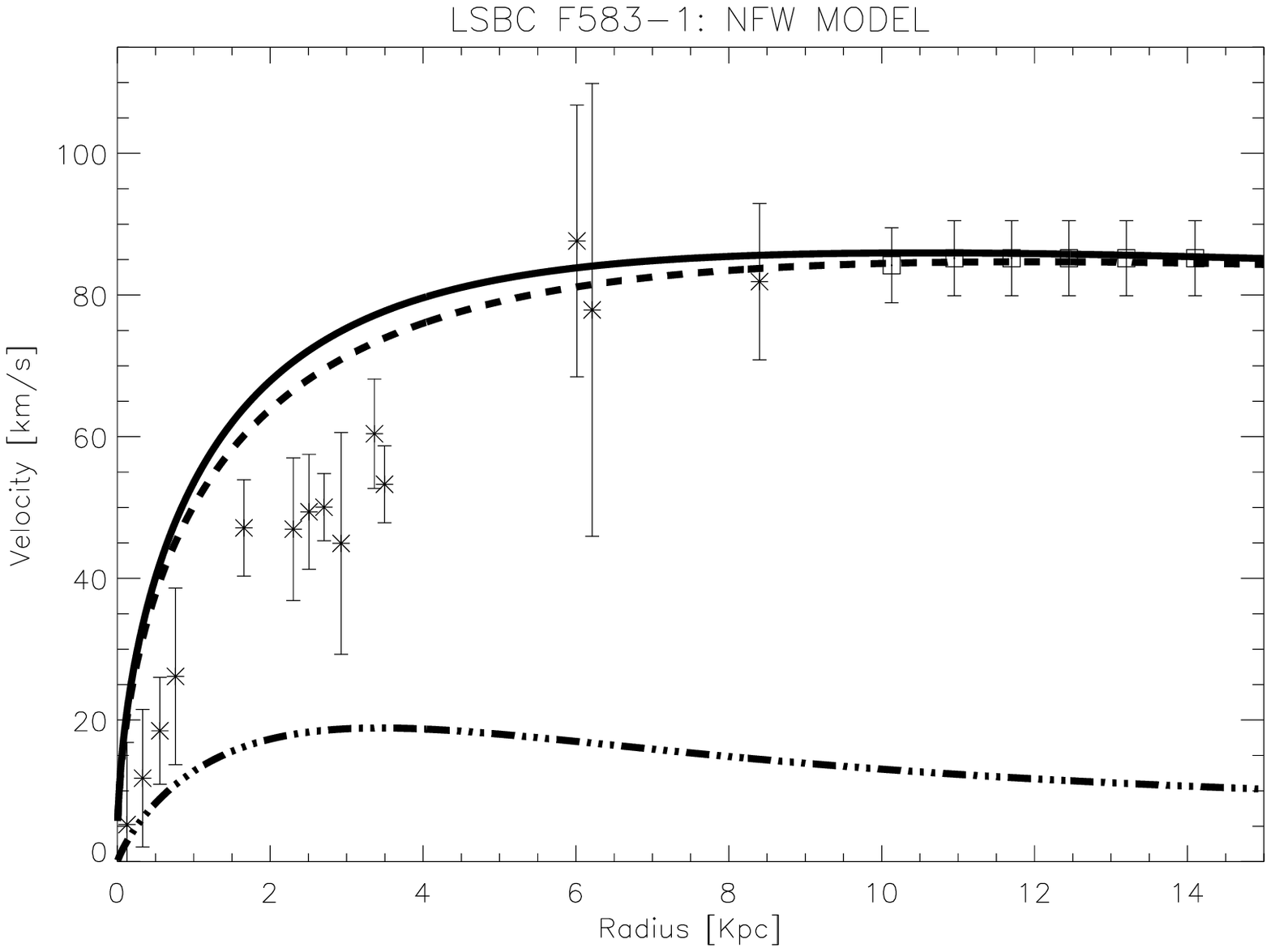}{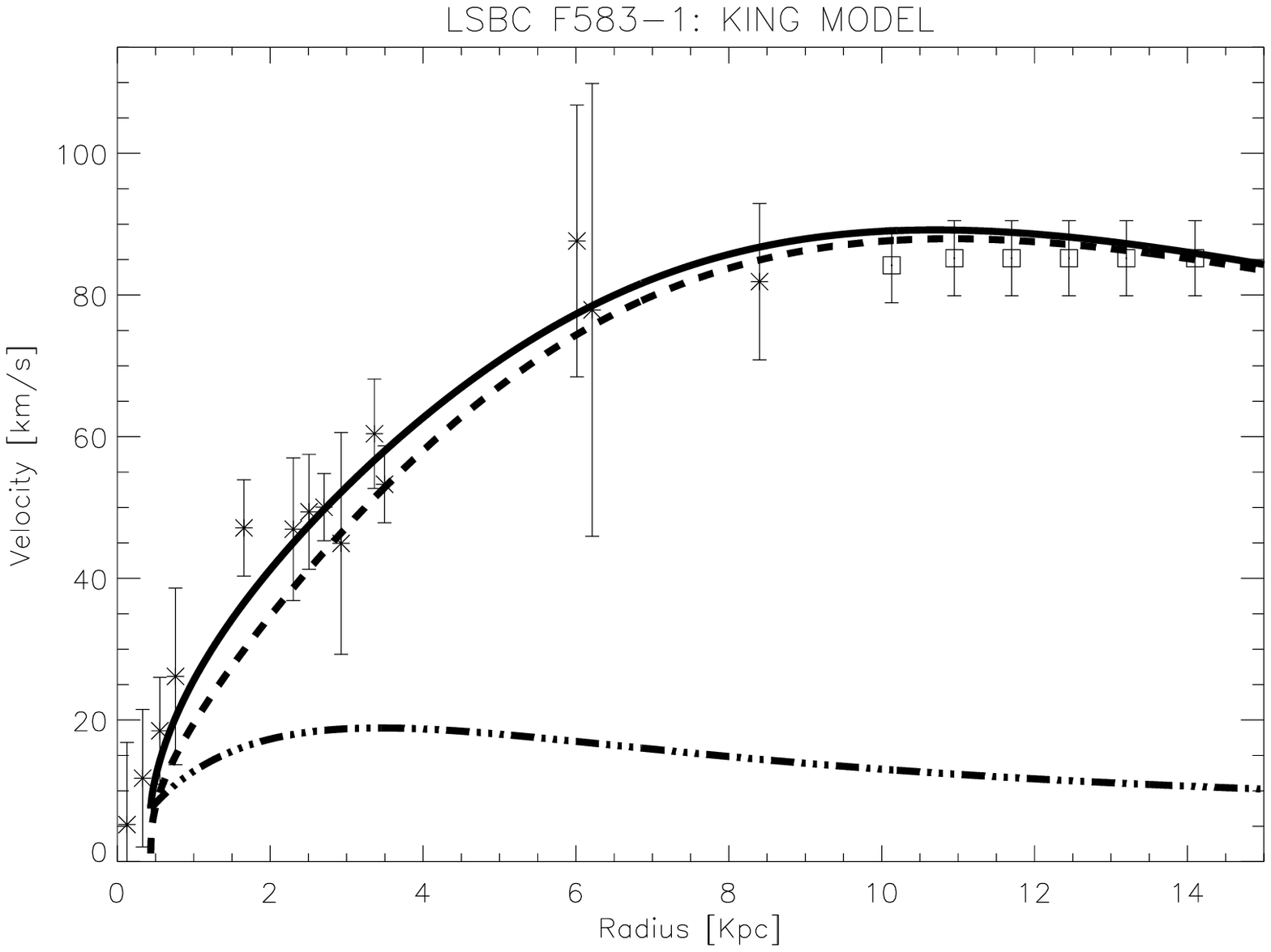}
\end{figure}

\clearpage

\end{document}